\documentclass[ALICE,manyauthors]{cernphprep}

\usepackage[comma,square,numbers,sort&compress]{natbib}
\usepackage{hyperref}

\usepackage{graphics}
\usepackage{graphicx}
\usepackage{amssymb}
\usepackage{subfigure}
\usepackage{marvosym}
\usepackage{amsmath}
\usepackage{multirow}
\usepackage{fancyhdr}
\usepackage{rotate}
\usepackage{lineno} 

\renewcommand{\d}{\mathrm{d}}
\newcommand{\pt}{p_\mathrm{T}}
\newcommand{\mt}{m_\mathrm{T}}
\newcommand{\ptmu}{p_{\mathrm{T},\mu}}
\newcommand{\snn}{\sqrt{s_\mathrm{NN}}}
\newcommand{\RpPb}{R_\textrm{pPb}}

\newcommand{\RFB}{R_\textrm{FB}}
\newcommand{\ALICE}{\mbox{ALICE }}

\begin{document}

\begin{titlepage}
\PHyear{2015}
\PHnumber{161}      
\PHdate{29 June}  
%

\title{$\pmb{\phi}$-meson production at forward rapidity in p-Pb collisions at $\pmb{\snn = 5.02}$~TeV and in pp collisions at $\pmb{\sqrt{s} = 2.76}$~TeV}
\ShortTitle{$\phi$-meson in p-Pb at $\snn = 5.02$~TeV and pp at $\sqrt{s}=2.76$~TeV}   

\Collaboration{ALICE Collaboration\thanks{See Appendix~\ref{app:collab} for the list of collaboration members}}
\ShortAuthor{ALICE Collaboration} 

\begin{abstract}
\noindent The first study of $\phi$-meson production in p-Pb collisions at forward and backward rapidity, at a
nucleon-nucleon centre-of-mass energy $\snn = 5.02$~TeV, has been
performed with the ALICE apparatus at the LHC.
The $\phi$-mesons have been identified in the dimuon decay channel in the
transverse momentum ($\pt$) range $1 < \pt < 7$~GeV/$c$, both 
in the p-going ($2.03 < y < 3.53$) and the Pb-going ($-4.46 < y < -2.96$)
directions --- where $y$ stands for the rapidity in the nucleon-nucleon centre-of-mass ---
the integrated luminosity amounting to $5.01 \pm 0.19$~nb$^{-1}$ and 
$5.81 \pm 0.20$~nb$^{-1}$, respectively, for the two data samples.
Differential cross sections as a function of transverse momentum and
rapidity are presented. The forward-backward ratio for 
$\phi$-meson production is measured for $2.96<|y|<3.53$, resulting in a
ratio $\sim 0.5$
with no significant $\pt$ dependence within the uncertainties.
The $\pt$ dependence of the $\phi$-meson nuclear modification factor $\RpPb$
exhibits an enhancement up to a factor~1.6 at $\pt = 3$-4~GeV/$c$ 
in the Pb-going direction.
The $\pt$ dependence of the $\phi$-meson cross section in pp
collisions at $\sqrt{s} = 2.76$~TeV, which is used to determine a
reference for the p-Pb results, is also presented here for $1 < \pt < 5$~GeV/$c$ and $2.5 <y < 4$, 
for a $78 \pm 3$~nb$^{-1}$ integrated luminosity sample.
\end{abstract}

\end{titlepage}
\setcounter{page}{2}


\section{Introduction}

\noindent Proton-nucleus (p-A) collisions are of special interest in the
context of high-energy nuclear physics for two reasons.
On one hand, a precise characterisation of particle production processes
in p-A collisions is needed as a reference for nucleus-nucleus
data. This allows in-medium effects --- linked to the formation of a
deconfined phase of the QCD matter, the quark-gluon
plasma~(QGP)~\cite{Bazavov:2011nk,Borsanyi:2010bp,Borsanyi:2010cj} ---
to be disentangled from the effects already present in cold nuclear matter. Among them, a sizeable role is played by the 
transverse momentum broadening of initial-state partons due to 
multiple scattering inside the nucleus, responsible for the Cronin
effect~\cite{Accardi:2002ik}, which may lead to an enhancement of
intermediate-$\pt$ hadron spectra.
In addition, p-A collisions at LHC energies provide a way to
probe the parton distributions of the
colliding nucleus at small values of Bjorken-$x$, 
in a regime where parton densities can reach saturation~\cite{Salgado:2011wc, Brandt:2014vva}.
In particular, the smallest~$x$ values contributing to
the wave function of the colliding nucleus can be probed 
by looking at particle production at large rapidities, 
in the p-going direction. 
Such a measurement can thus extend towards lower $x$-values the results of the lower-energy measurements by the PHOBOS and BRAHMS
experiments at RHIC~\cite{Arsene:2004fa, Back:2004je}.
Measurements of identified particle production may, in particular, provide useful
constraints for forthcoming theoretical studies of the saturation mechanism at small~$x$.

We have already reported results on charged particle production 
in p-Pb collisions at mid-rapidity. These results focused on
the pseudorapidity density~\cite{ALICE:2012xs} and the $\pt$ dependence of the nuclear modification 
factor~\cite{ALICE:2012mj,Abelev:2014dsa,Adam:2016dau}; the latter was found to be
consistent with unity for $\pt \gtrsim 2$~GeV/$c$. The nuclear modification
factor of charged hadrons was also studied by the BRAHMS and PHOBOS Collaborations
in \mbox{d-Au} collisions at the nucleon-nucleon centre-of-mass energy $\snn = 200$~GeV 
at RHIC~\cite{Arsene:2004ux, Back:2003hx},
as a function of pseudorapidity,  
where values smaller than unity were found for~$\eta \gtrsim 1$,
corresponding to the d-going direction.

In this Letter we report the measurement of $\phi$-meson production at forward
rapidity in p-Pb collisions at $\snn = 5.02$~TeV
in the transverse momentum ($\pt$) range $1 < \pt < 7$~GeV/$c$, for the centre-of-mass rapidity ($y$) ranges $2.03<y<3.53$ (p-going direction) and $-4.46<y<-2.96$
(Pb-going direction), in the dimuon decay channel with the 
\ALICE detector.
This measurement extends
the investigation of light-flavour particle production to forward 
rapidity. At the same time, it represents an essential baseline for the understanding of 
$\phi$ production in heavy-ion collisions, 
where an enhancement of strange particle yields relative to the ones measured in pp collisions 
has been proposed long ago as a signature of the
formation of a QGP phase~\cite{Rafelski:1982pu, Shor:1984ui,
  Koch:1986ud} triggering an intense experimental effort already at SPS and 
  RHIC energies~\cite{Alt:2004wc,Alt:2008iv,Alessandro:2003gy,Banicz:2009aa,Arnaldi:2011nn,Abelev:2008aa,Adler:2004hv}. 
It should be noted that, despite its hidden strangeness, producing a $\phi$-meson 
in a hadronic collision still implies the creation of a $s\bar{s}$ pair as it is the case for other strange hadrons, 
even if the hadronisation mechanisms can differ in reason of the different quark composition. 
In this context, the p-Pb data presented here
will provide an important reference for future measurements in Pb-Pb collisions 
in the LHC Run~2, which will be performed at a comparable energy.

The differential $\phi$-meson cross section as a function of transverse momentum is also presented
for pp collisions at $\sqrt{s} = 2.76$~TeV. This measurement complements the \ALICE results on
$\phi$-meson production in pp collisions at $\sqrt{s} = 7$~TeV, already reported in~\cite{ALICE:2011ad}
and, combined with the latter, is used to build the pp reference for the p-Pb measurements presented here.

\section{Experimental setup}

\noindent A full description of the \ALICE detector can be found
in~\cite{Aamodt:2008zz, Abelev:2014ffa}. 
The results presented in this Letter have been obtained detecting muon pairs 
with the muon spectrometer, covering 
the pseudorapidity region $-4 < \eta_\mathrm{lab} < -2.5$. Here and in
the following, the sign of $\eta_\mathrm{lab}$ is determined by the
choice of the LHC reference system. The other
detectors relevant for the analysis are the Silicon Pixel Detector~(SPD) 
of the Inner Tracking System~(ITS), the V0 detector and the Zero Degree Calorimeters~(ZDC).

The elements of the muon spectrometer are a hadron absorber, 
followed by a set of tracking stations, a dipole magnet, an iron wall acting
as muon filter and a set of trigger stations. 
The hadron absorber  is made of carbon, concrete and steel and is placed 
0.9~m away from the interaction point. 
Its total material budget corresponds to 10 hadronic interaction lengths. 
The dipole magnet provides an integrated magnetic field of 3~$\mathrm{T \cdot m}$ in the vertical direction. 
The muon tracking is provided by five tracking stations, each one composed of 
two cathode pad chambers. The first two stations are located 
upstream of the dipole magnet, the third one in the middle of its gap and the 
last two downstream of it.
A 1.2~m thick iron wall, corresponding to 7.2~hadronic interaction lengths, 
is placed between the tracking and trigger detectors and  
absorbs the residual secondary hadrons emerging from the hadron absorber. 
The hadron absorber together with the iron wall stops muons with total momentum lower than $\sim 4$~GeV/$c$.
The muon trigger detector consists of two stations, each one composed of two planes 
of resistive plate chambers, installed downstream of the muon filter. 
 
The SPD consists of two silicon pixel layers, covering the pseudorapidity regions 
$|{\eta_\mathrm{lab}}|<2.0$ and $|{\eta_\mathrm{lab}}|<1.4$ for the inner and outer layer, respectively. 
It is used for the determination of the primary interaction vertex position.
The V0 is composed of two scintillator hodoscopes 
covering the pseudorapidity regions 
$2.8< \eta_\mathrm{lab} <5.1$ and $-3.7< \eta_\mathrm{lab} <-1.7$.
It is used in the definition of the minimum
bias trigger signal, and allows the offline rejection of
beam-halo and beam-gas interactions to be performed. The ZDC
detectors, positioned symmetrically at 112.5~m from the interaction point, are used to clean the event
sample by removing beam-beam collisions not originating from nominal LHC bunches.

\section{Data selection and signal extraction}

\noindent The analysis presented in this Letter is based on two data samples, collected by
\ALICE during the 2013 p-Pb and pp runs at $\snn = 5.02$~TeV and $\sqrt{s} = 2.76$~TeV, respectively.
In this section we present the details of the data selection, 
as well as the procedure followed for the extraction of the $\phi$-meson signal.

\subsection{Data selection}

\noindent The Minimum-Bias~(MB) trigger for the considered data sample
is given by the logical AND of the signals in the two V0 detectors~\cite{Gagliardi:2011he}.
Events containing a muon pair are selected by means of
a specific dimuon trigger, based on the 
detection of two muon candidate tracks in the trigger system of the
muon spectrometer, in coincidence with the MB condition.
Due to the intrinsic momentum cut imposed by the detector, only muons with $\pt \gtrsim 0.5$~GeV/$c$
manage to leave a signal in the trigger chambers.

Because of the different energy of the LHC proton and Pb beams ($E_\mathrm{p} = 4$~TeV, 
$E_\mathrm{Pb} = 1.58$~A$\cdot$TeV), in p-Pb collisions the nucleon-nucleon centre-of-mass moves in the laboratory
with a rapidity $y_0 = 0.465$
in the direction of the proton beam. 
The directions of the proton and Pb beam orbits were inverted during the p-Pb data taking period.
This allowed the \ALICE muon spectrometer to access two different rapidity regions\footnote{~The sign of $y$ is
defined by assuming the proton beam to have positive rapidity.}: the region $2.03 <y < 3.53$ where the proton 
beam is directed towards the muon spectrometer
(p-going direction) and the region $-4.46 < y < -2.96$ where the
Pb beam is directed towards the muon spectrometer (Pb-going direction).
In the following, these two rapidity ranges are also referred to as
``forward'' and ``backward'', respectively. For pp collisions at $\sqrt{s} = 2.76$~TeV the muon spectrometer
covers the rapidity region $2.5 <y < 4$\footnote{~In this case the
  sign of $y$ is defined by assuming the proton beam entering the muon
  spectrometer to have positive rapidity.}. 

Background events not coming from beam-beam interactions are rejected
by performing an offline selection,
based on the requirement that the timing signals from the V0 and ZDC detectors are compatible with
a collision occurring in the fiducial interaction region
$|z_\mathrm{vtx}| \lesssim 10$~cm.

The integrated luminosity for the p-Pb data samples was evaluated as
$L_\mathrm{int} = N_\mathrm{MB}/\sigma_\mathrm{MB}$, where
$N_\mathrm{MB}$ is the number of MB events corresponding to the analysed
triggered events, and $\sigma_\mathrm{MB}$ the MB trigger cross
section. The value of $N_\mathrm{MB}$ was obtained by averaging the results
of two different methods --- one based on the ratio of trigger rates and
the other based on the offline selection of dimuon events in the MB
data sample~\cite{Abelev:2013yxa} ---
while the MB trigger cross sections $\sigma_{\rm{MB}}$ were measured with a
van der Meer scan and found to be $2.09 \pm 0.07$~b and $2.12 \pm
0.07$~b, respectively, for the beam configurations corresponding to the forward and
backward rapidity coverage of the muon
spectrometer~\cite{Abelev:2014epa}.
For the pp data sample, the integrated luminosity is calculated with
the method described in~\cite{Abelev:2014qha}, using as reference the
MB trigger cross section $\sigma_\mathrm{MB} = 47.7 \pm 0.9$~mb,
measured in a van der Meer
scan~\cite{Abelev:2012sea}.

The resulting values of $L_\mathrm{int}$ for the analysed p-Pb data samples are $5.01
\pm 0.19$~nb$^{-1}$ and $5.81 \pm
0.20$~nb$^{-1}$~\cite{Abelev:2014epa, Abelev:2013yxa} ---
corresponding to $\sim 24\,000$ and $\sim 26\,000$ reconstructed $\phi
\to \mu\mu$ decays (see next section) ---
respectively for the forward and backward rapidity regions.
For the pp data sample, the integrated luminosity amounts to $78 \pm 3$~nb$^{-1}$
for a total number of $\sim 1\,400$ reconstructed $\phi
\to \mu\mu$ decays.

Track reconstruction in the muon spectrometer is based on a Kalman filter algorithm~\cite{ALICE:2011ad,OfflineNote,kalmanC}.
Muon identification is performed by requiring the candidate track to match a track segment 
in the trigger chambers (trigger tracklet). This request selects muons
with $\ptmu \gtrsim 0.5$~GeV/$c$
and, as a consequence, significantly affects the collected statistics
for dimuons with invariant mass $\lesssim 1$~GeV/$c^2$ and $\pt \lesssim 1$~GeV/$c$. 
It is also required that muon 
tracks lie in the pseudorapidity interval $-4 < \eta_{\mu} < -2.5$, where $\eta_{\mu}$ is defined in the laboratory frame, 
in order to remove the tracks close to the acceptance borders
of the spectrometer, where the acceptance drops abruptly. 
Selected tracks are finally required to exit the hadron absorber at a
radial distance from the beam axis, $R_\mathrm{abs}$, in the range $17.6
< R_\mathrm{abs} < 89.5$~cm: this cut, for all practical purposes equivalent to
the one on $\eta_\mu$, explicitly ensures the rejection of tracks crossing the region of
the absorber with the highest density material, where multiple scattering
and energy loss effects are large and can affect the mass resolution.
Muon pairs are built combining two muon tracks that satisfy the above cuts.

\subsection{Signal extraction}

\noindent The Opposite-Sign (OS) muon pairs are composed of correlated and uncorrelated pairs. 
The former contain the signal of interest for the present analysis, while the latter --- 
mainly coming from semi-muonic decays of pions and kaons --- form a
combinatorial background. The contribution of the combinatorial
background to the OS mass spectrum
 was evaluated using an event mixing technique in which uncorrelated pairs are formed with muons taken from different events.
A detailed description of the technique can be 
found in~\cite{ALICE:2011ad}. 
The ratio between correlated and uncorrelated OS dimuons at the $\phi$-meson mass
is $\sim 0.65$ ($\sim 0.40$) in p-Pb collisions at $\snn = 5.02$~TeV at forward (backward) rapidity,
and $\sim 1.30$ in pp collisions at $\sqrt{s} = 2.76$~TeV. A direct comparison of the raw OS mass spectrum and
the associated combinatorial background is presented in~\cite{ALICE-PUBLIC-2016-004}, for each of the $\pt$ intervals considered
in the analysis.

\begin{figure}[p] 
   \begin{center}
   \vspace{-0.7cm}
     \includegraphics[width=0.48\textwidth]{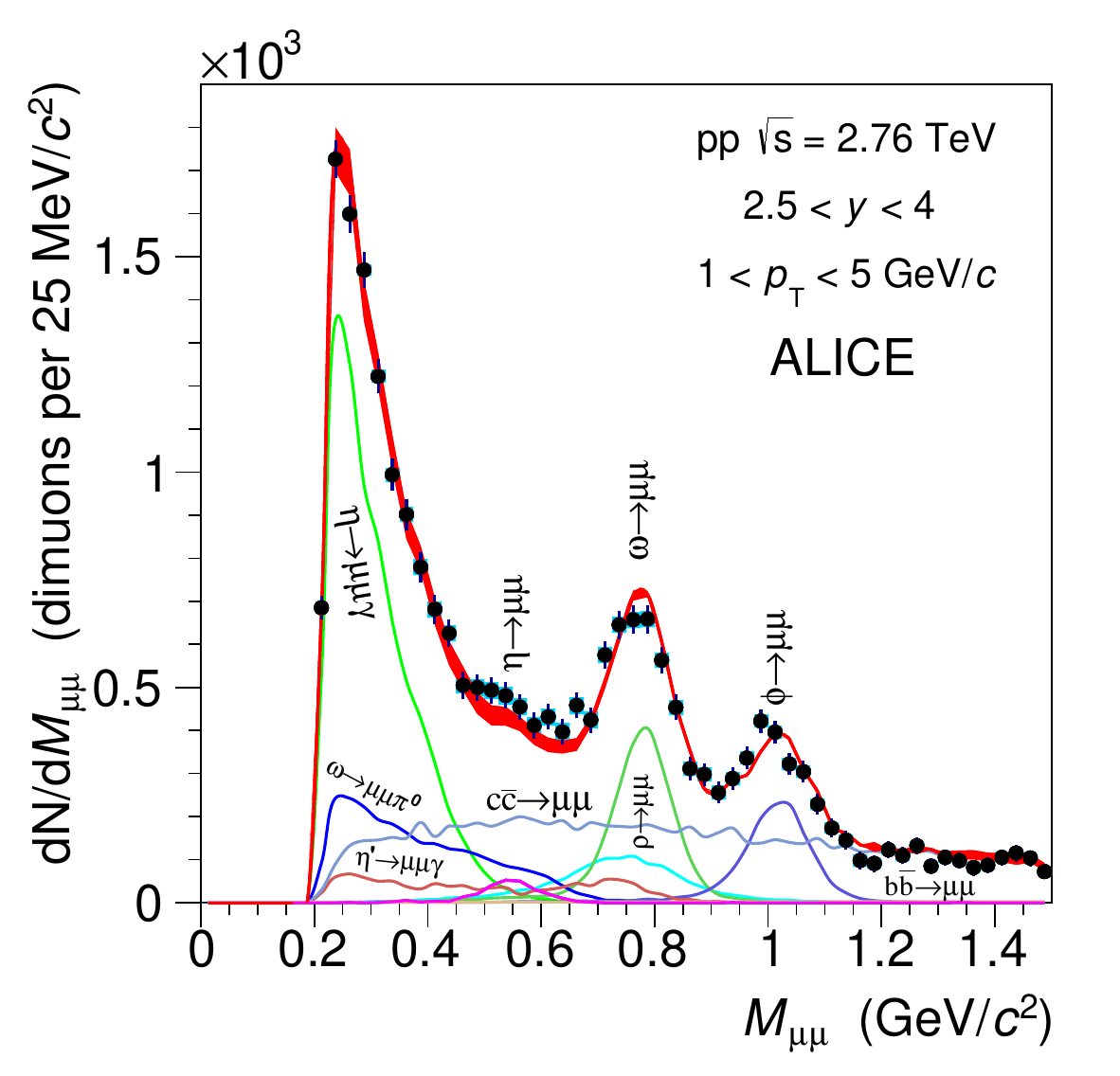}
     \includegraphics[width=0.48\textwidth]{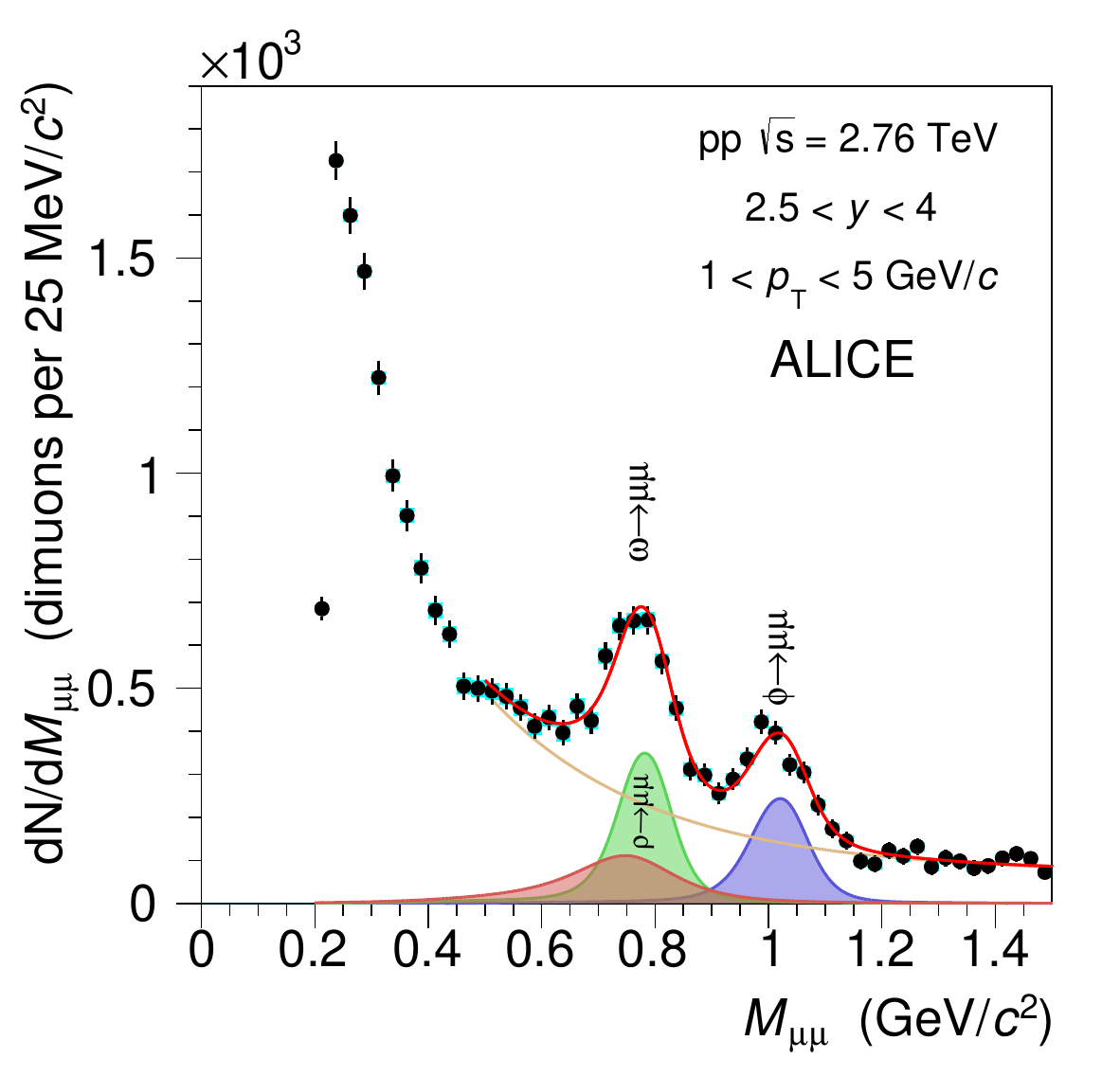} \\
   \vspace{-0.7cm}
     \includegraphics[width=0.48\textwidth]{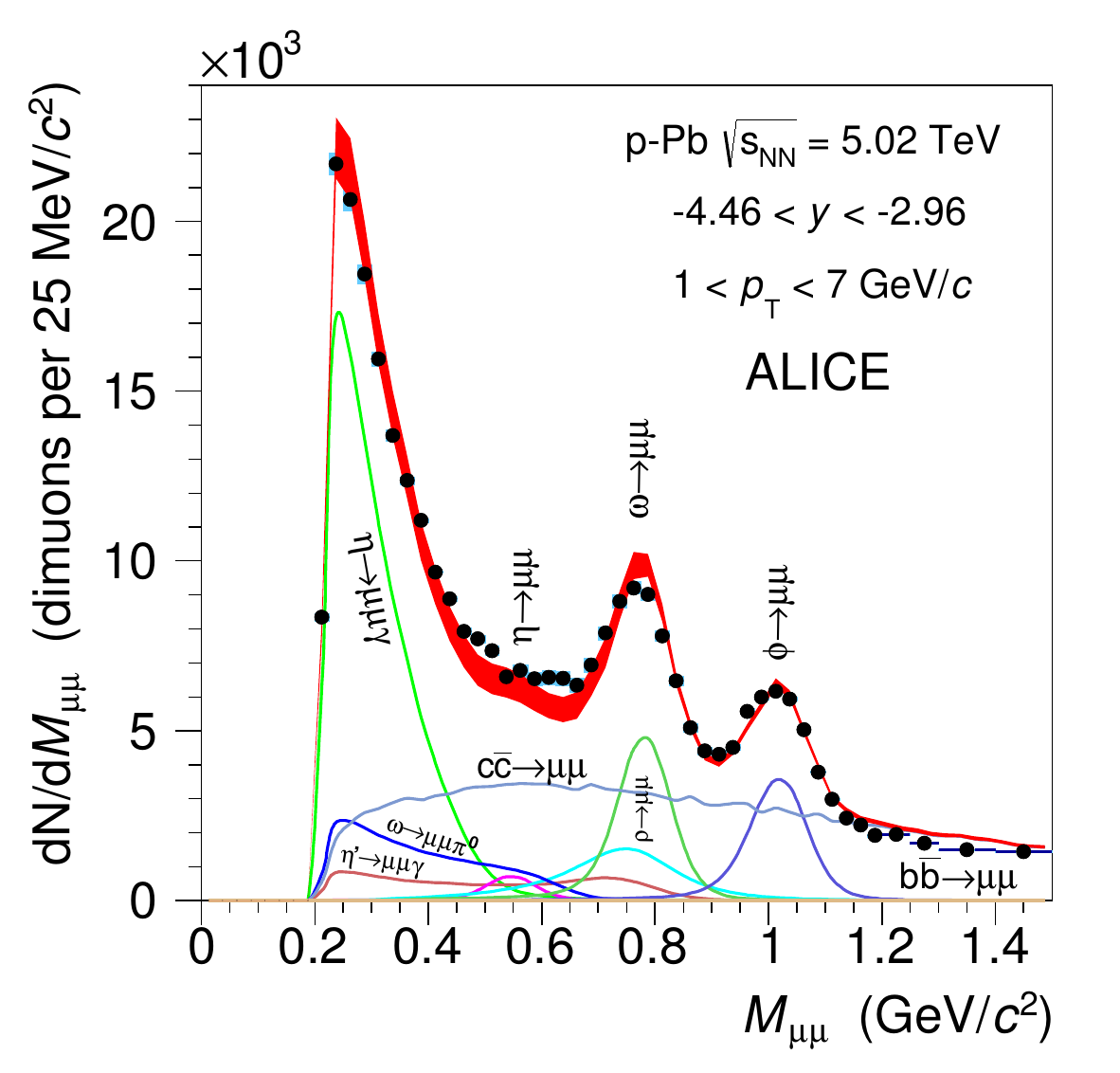}
     \includegraphics[width=0.48\textwidth]{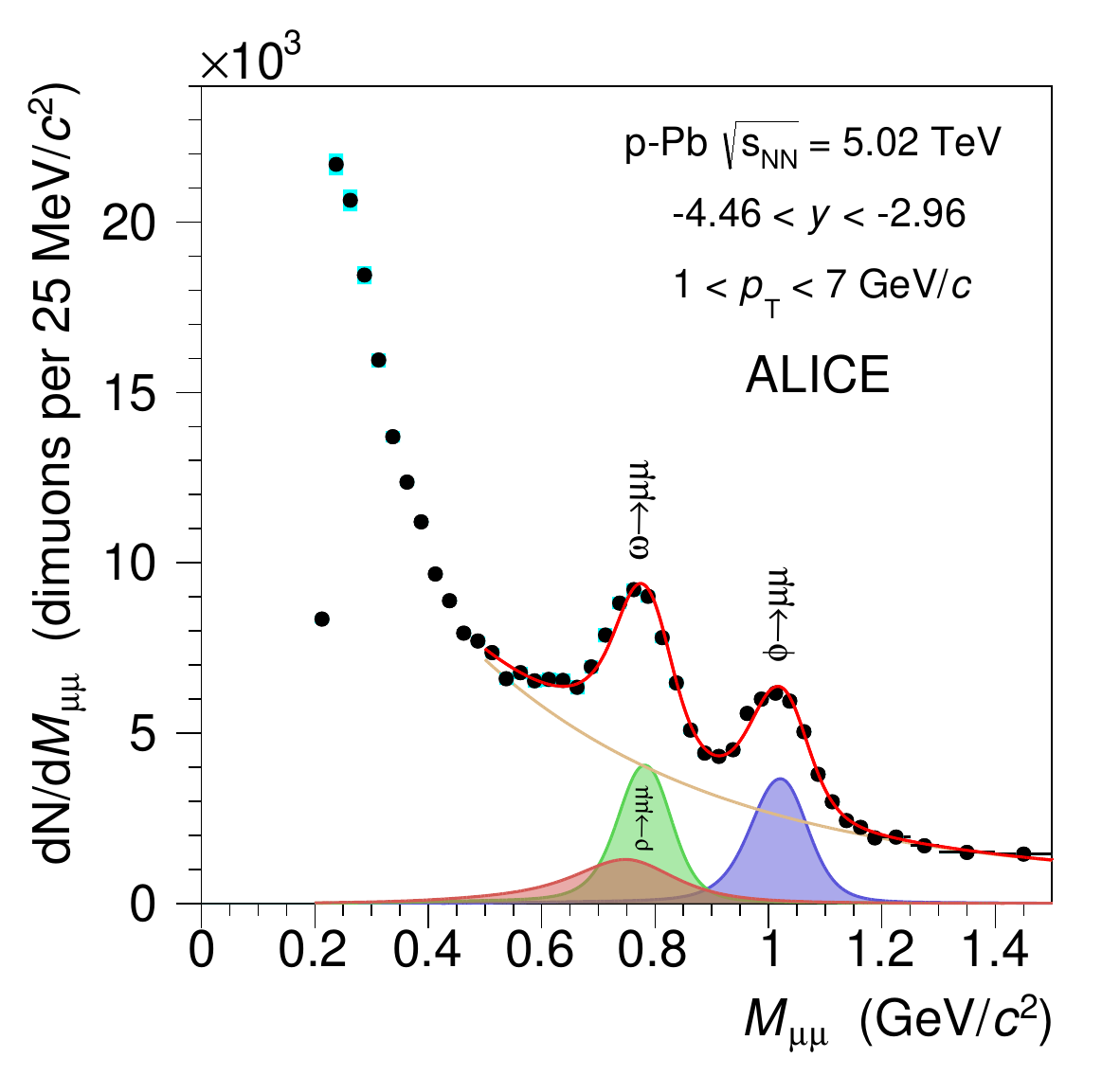} \\
   \vspace{-0.7cm}
     \includegraphics[width=0.48\textwidth]{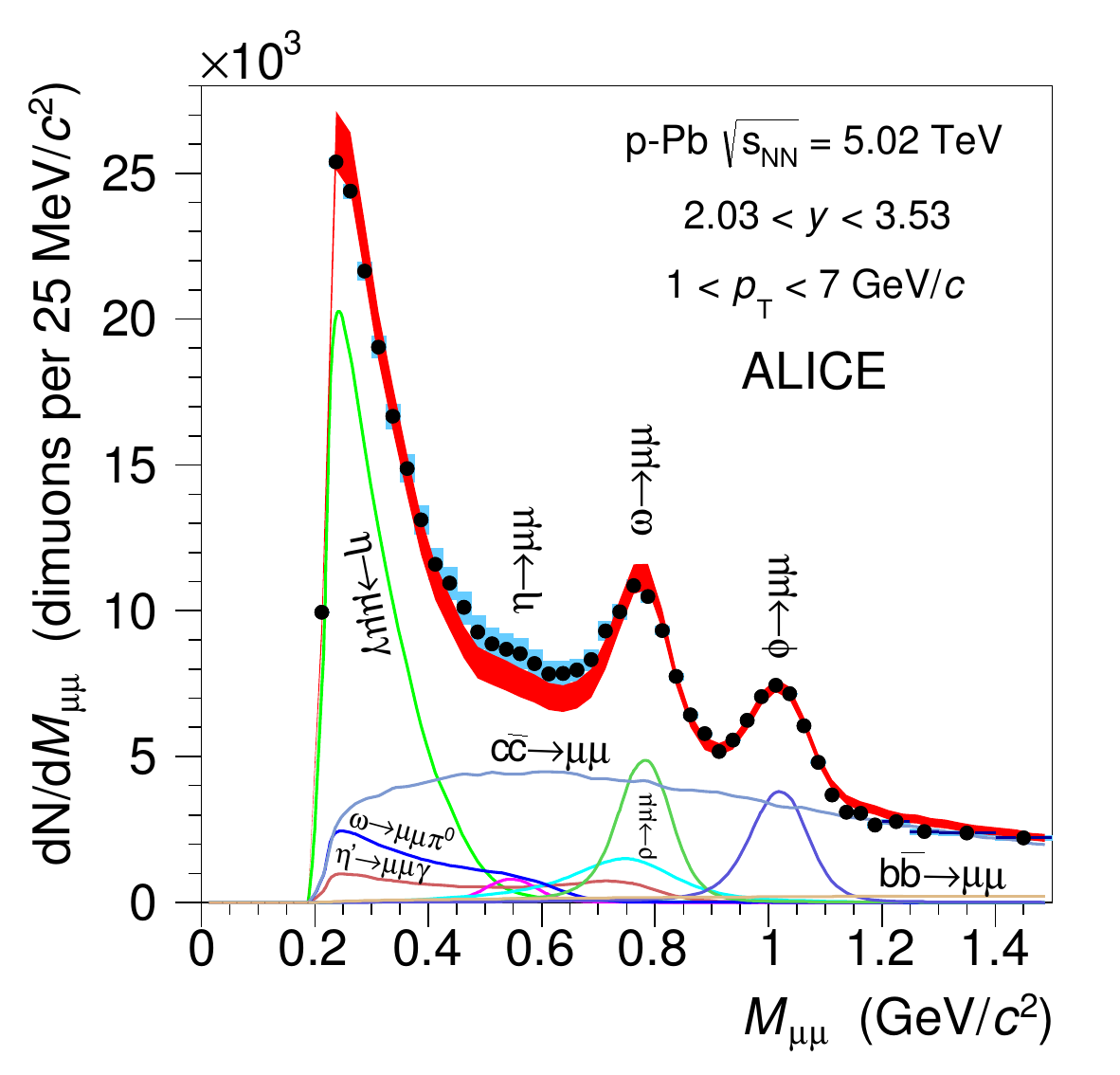}
     \includegraphics[width=0.48\textwidth]{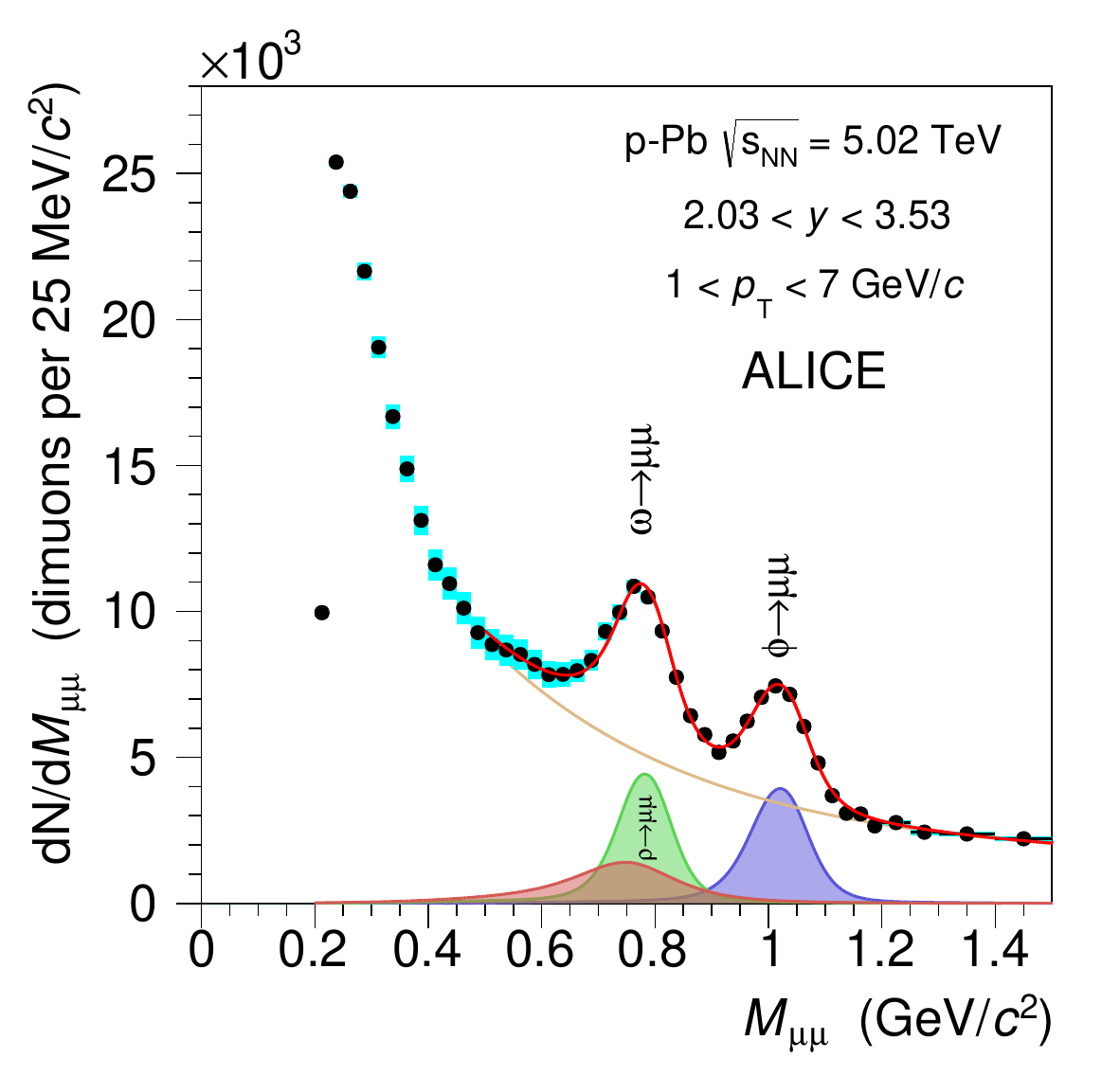}
   \end{center} 
   \vspace{-0.4cm}
   \caption[\textwidth]{Dimuon mass spectrum after combinatorial background subtraction: 
     $\pt$-integrated pp sample (top panels) and $\pt$-integrated p-Pb
     sample in the backward (centre panels) and forward (bottom panels) 
     rapidity regions, compared to the result of the hadronic-cocktail and the empirical-function fits 
     (left- and right-column panels, respectively). Error boxes on data points
     (well visible only in some regions on the plots) represent the systematic 
     uncertainty due to the combinatorial background subtraction, while error bars 
     account for the statistical uncertainty. 
     The width of the hadronic-cocktail fit result (red band) combines the statistical 
     uncertainties of the free fit parameters with the systematic uncertainties on the fixed parameters (see text).}
   \label{fig:fitMassSpectrum}
   \vspace{0.2cm}
 \end{figure}

The invariant mass spectra in pp and p-Pb collisions, obtained after combinatorial background subtraction, 
are shown in \figurename~\ref{fig:fitMassSpectrum} for the $\pt$-integrated samples. In the 
left-column panels of \figurename~\ref{fig:fitMassSpectrum}, the signal is described 
in the low-mass region (from the threshold up to $\sim 1.5$~GeV/$c^2$)
by the superposition of 
a so-called hadronic cocktail and the open charm and open beauty processes. 
The processes included in the hadronic cocktail are the two-body and Dalitz decays of the light neutral mesons 
$\eta, \rho, \omega, \eta'$ and~$\phi$, which dominate dimuon production 
for invariant masses below $\sim 1$~GeV/$c^2$. 
The open charm and open beauty contributions
arise from correlated semi-muonic decays of charm and beauty mesons
and baryons.

The hadronic cocktail was simulated with a dedicated generator described in~\cite{ALICE:2011ad},
tuned to the existing measurements whenever possible, otherwise based on the kinematic distributions extracted from \texttt{PYTHIA}~\cite{Sjostrand:2006za}.
In particular, the kinematic distributions of the $\phi$-meson have been tuned 
by means of an iterative procedure to the results presented in this Letter to ensure self-consistency for this analysis.
The open charm and beauty generation 
is based on a parameterisation of the spectra generated with \texttt{PYTHIA}~\cite{OfflineNote}. The detector response 
for all these processes is obtained with a simulation based on the \texttt{GEANT3}~\cite{Brun:1994aa} transport code. 
Simulated events are then subjected to the same reconstruction 
and selection procedure as the data. 

When describing the signal with the superposition of the 
aforementioned contributions, four parameters are adjusted in the fit procedure in each of the $\pt$
or rapidity intervals considered in the analysis: the yield of the $\eta$, $\omega$
and $\phi$-mesons, and the one of the open charm
and beauty processes, with the relative beauty/charm contribution fixed (see later in this paragraph). In this way, each parameter is
linked to a process dominating in at least one region of the considered mass spectrum.
The remaining degrees of freedom are fixed either according to the 
relative branching ratios known from literature~\cite{Beringer:1900zz}, or assuming specific hypotheses on the cross section ratios.
In particular, the production cross section of the $\rho$-meson is assumed to be
the same as for the $\omega$ as suggested from both models and pp data~\cite{ALICE:2011ad}, while the $\eta'$ contribution was 
derived from the $\eta$ cross section by applying the ratio of the
corresponding cross sections $\sigma_{\eta'}/\sigma_\eta = 0.3$ taken
from the \texttt{PYTHIA} tunes \texttt{ATLAS-CSC} and \texttt{D6T}
which best describe the available low-mass dimuon measurements at the
LHC energies~\cite{ALICE:2011ad}. 
The open beauty normalisation is fixed to the open charm one via a fit of
the $\pt$- and rapidity-integrated mass spectra 
in which the yields from both processes are free parameters;
when performing differential studies, the beauty/charm ratio is scaled
according to the differential distributions for the two processes, given by the 
Monte Carlo (MC) simulations.

For each $\pt$ and rapidity interval, the raw number of 
$\phi$-mesons is determined via a fit procedure based on a $\chi^2$ minimisation, performed on the signal 
obtained after the subtraction of the combinatorial background,
shown in~\figurename~\ref{fig:fitMassSpectrum} for the
$\pt$-integrated samples. Several tests have been performed to evaluate the robustness of the 
signal extraction and estimate an appropriate systematic uncertainty for it. They include in particular:
\begin{itemize}
 \item Replacing the fit based on the full MC hadronic cocktail with a fit
   based on the superposition of 
   various empirical functions. In this case, illustrated in the right-column panels of \figurename~\ref{fig:fitMassSpectrum}, the continuum is modelled either with exponential functions
   or variable-width Gaussians, while the $\rho$+$\omega$ and $\phi$-meson peaks are described by 
   Crystal Ball functions~\cite{Oreglia:1980cs} tuned on the MC.
 \item Varying the ratio between the yields of open beauty and open
   charm processes. It was verified that for perturbations as large as
   $\pm 50\,\%$ (resulting in a reasonably wide range of variation for the
   shape of the total continuum) no significant systematic effect
   is visible.
 \item Varying the ratios between the two-body and Dalitz branching ratios of the $\eta$ and $\omega$-mesons, as well as the
   cross section ratios $\sigma_\rho/\sigma_\omega$ and $\sigma_{\eta'}/\sigma_\eta$, within the uncertainties 
   coming either from the available measurements or from the differences between the \texttt{PYTHIA} 
   tunes considered in the analysis of the pp data. The branching ratio 
   $BR_{\omega\to\mu\mu}$ was taken
   as the average (weighted by the corresponding uncertainties) of the available measurements of
   $BR_{\omega\to\mu\mu}$ and $BR_{\omega\to ee}$~ \cite{Beringer:1900zz}, assuming lepton universality. 
 \item Varying the considered fit range: in particular, the fit was performed both including and excluding 
   the mass region from 0.4 to 0.65~GeV/$c^2$ where the quality of
   the comparison between the data and the sum of the MC sources turns out to be lower.
 \end{itemize}
The total systematic uncertainty on the signal extraction was taken as the quadratic sum of the above sources. 
The systematic uncertainty on the combinatorial background is estimated by comparing the shape of the Like-Sign
dimuon contributions coming from the event mixing procedure and from the raw data~\cite{ALICE:2011ad}. This 
uncertainty depends on the mass, 
its relative contribution being maximal in the mass window 0.5-0.8~GeV$/c^2$ and minimal around the $\phi$-meson peak, and
it is added in quadrature, for each point of the mass spectrum, to the statistical uncertainty of the signal: in this way, this source of systematics 
is accounted for by the $\chi^2$ minimisation procedure, and automatically propagated when evaluating the $\phi$-meson raw signal 
from the fit parameters.  The uncertainty associated to the sum of the MC sources
 (red band in the left-column plots of \figurename~\ref{fig:fitMassSpectrum})
 is evaluated by combining the uncertainties on the normalisation of
 each considered process. For the processes whose normalisation is
 left free in the fit, this uncertainty is the statistical one resulting
 from the fit procedure itself; for the rest of the processes, we also propagate the
 systematic uncertainty on the parameters (branching ratios or cross
 section ratios) which fix their normalisations to those of the free processes.
 
 \section{Results}
 
 \noindent The results of the $\phi$-meson analysis are presented as follows.
 We first present the measurement of the production cross sections,
 starting with its $\pt$-dependence in pp collisions at $\sqrt{s} =
 2.76$~TeV, followed by p-Pb collision results 
 as a function of $\pt$ and rapidity. Then, we show the ratio of the cross sections measured in the forward and backward regions, 
 obtained in the common rapidity interval $2.96 < |y| < 3.53$.
 Finally, the measurement of the nuclear modification factor $\RpPb$ as a function of $\pt$ is presented, separately for
 the p-going and the Pb-going directions.
 
 \subsection{Production cross section in pp and p-Pb collisions}

\noindent The cross section $\sigma_\phi$ was evaluated for each $\pt$ and rapidity interval as: 
\begin{displaymath}
  \sigma_\phi(x) = \frac{N^{\mathrm{raw}}_{\phi\to\mu\mu}(x)}
    {[A\cdot\epsilon](x)\cdot BR_{\phi\to\mu\mu}\cdot L_\mathrm{int}}~, 
\end{displaymath}
where $x$ stands for any specific $\pt$ or rapidity interval
considered.
The total systematic uncertainty on $N^{\mathrm{raw}}_{\phi\to\mu\mu}(x)$, after combining the different sources described above,
ranges between 3\,\% and 8\,\% depending on the collision system and kinematic range. The branching ratio 
$BR_{\phi\to\mu\mu}$ was taken from~\cite{Beringer:1900zz}
as the average (weighted by the corresponding uncertainties) of the available measurements of
$BR_{\phi\to\mu\mu}$ and $BR_{\phi\to ee}$, assuming lepton universality, resulting in a final uncertainty of 
approximately~1\,\%. 
The product of the geometrical acceptance~$A$ and the reconstruction efficiency~$\epsilon$ has been evaluated
by means of MC simulations, using the cocktail predictions
for the differential input spectra. The values are obtained  as the
ratio between the
number of dimuons at the output of the reconstruction chain ---
including the effect of the event selection criteria imposed on the data --- and the
number of dimuons injected as input. 

The uncertainty on $[A\cdot\epsilon]$ mainly
originates from the systematic uncertainty on the dimuon tracking and trigger
efficiencies. The systematic uncertainty on the tracking efficiency, amounting to 6\,\% and 4\,\% 
for the backward and forward rapidity regions, respectively, comes from the residual differences between 
the results of the efficiency-determination method based on reconstructed tracks~\cite{Aamodt:2011gj,Abelev:2013yxa},
applied to both data and MC. For the systematic uncertainty on the trigger efficiency, we also refer to
the procedure discussed in~\cite{Abelev:2013yxa}, resulting in an uncertainty of 3.2\,\% and 2.8\,\%, respectively, 
for the backward and forward rapidity regions considered in the analysis. In order to test possible 
additional systematic effects related to the hardware
trigger $\pt$ cut, imposing a non-sharp threshold around 0.5~GeV/$c$, 
the analysis was repeated with the additional offline sharp cuts $\ptmu > 0.5$~GeV/$c$ 
and $\ptmu > 1$~GeV/$c$ on single muons. For each of the two
alternative scenarios, the corresponding measurement of the 
$\phi$-meson cross section was compared to the one coming from the reference
analysis: the difference between the results was found to be
smaller than the quadratic difference of the statistical
uncertainties, showing that no significant bias related to the trigger threshold
affects the results~\cite{Barlow:2002yb}.

The reported values correspond to a zero-polarisation scenario for the 2-body decay of the $\phi$-meson,
in the absence of evidence supporting less trivial assumptions (in particular, no measurement of $\phi$-meson polarisation 
is currently available at the LHC energies).


\subsubsection{Production cross section in pp collisions}
\label{cross_section_pp}

\begin{figure}[b!] 
   \begin{center}
    \includegraphics[width=0.60\textwidth]{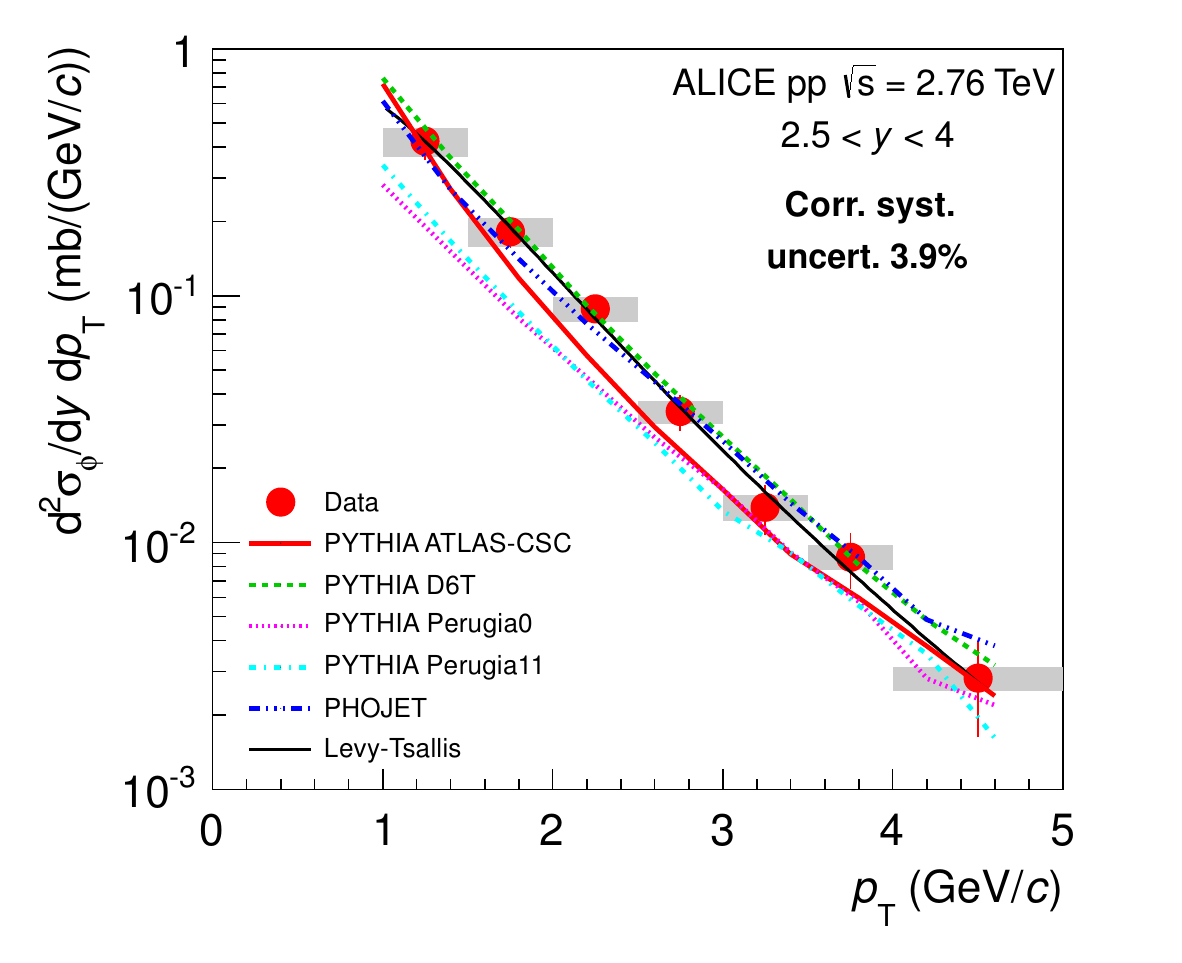}
    \end{center} 
    \vspace{-0.4cm}
    \caption[\textwidth]{$\phi$-meson cross section as a function of $\pt$ in pp collisions at $\sqrt{s} = 2.76$~TeV.
    Error bars and boxes represent statistical and systematic uncertainties, respectively.
    Predictions from \texttt{PHOJET}~\cite{Engel:1995yda} and the \texttt{PYTHIA} tunes \texttt{ATLAS-CSC}~\cite{Buttar:2004iy}, \texttt{D6T}~\cite{Field:2008zz}, 
    \texttt{Perugia0} and \texttt{Perugia11}~\cite{Skands:2010ak} are also shown for comparison, as well as the result of a fit 
    with the Levy-Tsallis function defined by Eq.~(\ref{eqn:ptLevy}).}
\label{fig:phiYieldVsPt_pp}
\vspace{0.1cm}
\end{figure}

The inclusive, $\pt$-differential $\phi$-meson cross section in pp collisions at $\sqrt{s} = 2.76$~TeV is shown in
\figurename~\ref{fig:phiYieldVsPt_pp}. 
The data points, also summarised in \tablename~\ref{tab:ppCrossSection_2.76}, 
are compared with the predictions from \texttt{PHOJET}~\cite{Engel:1995yda}
and \texttt{PYTHIA}~\cite{Sjostrand:2006za}, where for the latter the \texttt{Perugia0}, \texttt{Perugia11}~\cite{Skands:2010ak},
\texttt{ATLAS-CSC}~\cite{Buttar:2004iy}, and
\texttt{D6T}~\cite{Field:2008zz} tunes are considered. 
An~overall good agreement is found between predictions and data, with the exception of the \texttt{Perugia0} and
\texttt{Perugia11} tunes of \texttt{PYTHIA} which underestimate the measured cross section by a factor of two, as already
observed for the $\phi$-meson measurements at $\sqrt{s} =
7$~TeV~\cite{ALICE:2011ad,Abelev:2012hy}. It is worth to note that the
\texttt{D6T} tune is not successful in describing the $\pt$ evolution
of the $K/\pi$ ratio at mid-rapidity in pp collisions at $\sqrt{s} =
2.76$~TeV, as measured by the CMS
Collaboration~\cite{Chatrchyan:2012qb}: this suggests that hidden
strangeness is better reproduced than open strangeness in this
specific \texttt{PYTHIA} tune.
Data points were fitted with a Levy-Tsallis function~\cite{Tsallis:1987eu}
\begin{equation}
 \frac{1}{\pt} \frac{\d N}{\d \pt} \propto \left( 1 + \frac{\mt - m_\phi}{n T} \right)^{-n}~,
 \label{eqn:ptLevy}
\end{equation}
where $\mt = \sqrt{\pt^2 + m_\phi^2}$ stands for the transverse mass, 
obtaining the values $n = 10.2 \pm 4.8$ and $T = 284 \pm 72$~MeV for
the fit parameters,
where the errors reflect the statistical uncertainties only.
The cross section integrated over the accessible $\pt$ range $1<\pt<5~\mathrm{GeV}/c$ is 
$\sigma_\phi = 0.566 \pm
0.055~\mathrm{(stat.)} \pm 0.044~\mathrm{(syst.)}$~mb.
The systematic uncertainties for this measurement are summarised in \tablename~\ref{tab:systematics_pp}.

\begin{table}[ht!]
  \begin{center}
    \begin{tabular}{l c l}
    \hline \hline
    $\pt$~(GeV/$c$)~~~~~~~~~~ & ~~$\chi^2$/ndf~~ & $\d^2 \sigma_\phi / (\d y \d \pt)$~(mb/(GeV/$c$)) \\ 
    \hline \hline
     $[1.0, 1.5]$ & 1.1 & $0.423   \pm 0.067  \pm 0.043 $  \\
     $[1.5, 2.0]$ & 1.7 & $0.182   \pm 0.025  \pm 0.018 $  \\
     $[2.0, 2.5]$ & 1.1 & $0.089   \pm 0.011  \pm 0.007 $  \\
     $[2.5, 3.0]$ & 1.1 & $0.0340  \pm 0.0056 \pm 0.0020$  \\
     $[3.0, 3.5]$ & 0.9 & $0.0139  \pm 0.0032 \pm 0.0011$  \\
     $[3.5, 4.0]$ & 1.1 & $0.0087  \pm 0.0022 \pm 0.0006$  \\
     $[4.0, 5.0]$ & 1.1 & $0.0028  \pm 0.0012 \pm 0.0002$  \\
    \hline \hline
    \end{tabular}
  \caption{$\pt$-differential production cross section for the $\phi$-meson in pp collisions at $\sqrt{s} = 2.76$~TeV, for $2.5 < y < 4$. 
  The first uncertainty is statistical and the second is the bin-to-bin uncorrelated systematic. The bin-to-bin correlated relative systematic uncertainty 
  is 3.9\,\%. The $\chi^2$/ndf values are relative to the hadronic-cocktail fit and the
  [0.8,~1.2~GeV/$c^2$] mass region, where $\mathrm{ndf} = 10$.}
  \label{tab:ppCrossSection_2.76}
  \end{center}
\end{table}

\begin{table}[ht!]
  \begin{center}
    \begin{tabular}{l c }
    \hline \hline
    Source~~~~~~~~~~~~~~~~~~~~~~~~~~~ & Syst.~uncertainty on $\sigma^\textrm{pp}_\phi$ \\ 
    \hline \hline
    \emph{Uncorrelated} &  \\
    \hline
    Signal extraction                &     3--8\,\%  \\
    Tracking efficiency              &     4\,\%    \\ 
    Trigger efficiency               &     3\,\%   \\ 
    \hline \hline
    \emph{Correlated} &  \\
    \hline
     $L_\mathrm{int}$                &     3.8\,\% \\
     $BR(\phi \to \ell\ell$)          &     1\,\%   \\ 
    \hline \hline
    \end{tabular}
  \caption{Systematic uncertainties (in percent) contributing to the
    measurement of the $\phi$-meson cross section in pp collisions at 
    $\sqrt{s} = 2.76$~TeV. When the uncertainty values depend on the $\pt$
    interval, their minimum and maximum values are quoted.}
  \label{tab:systematics_pp}
  \end{center}
\end{table}

\subsubsection{Production cross section in p-Pb collisions}

\begin{figure}[t!] 
   \begin{center}
    \includegraphics[width=0.49\textwidth]{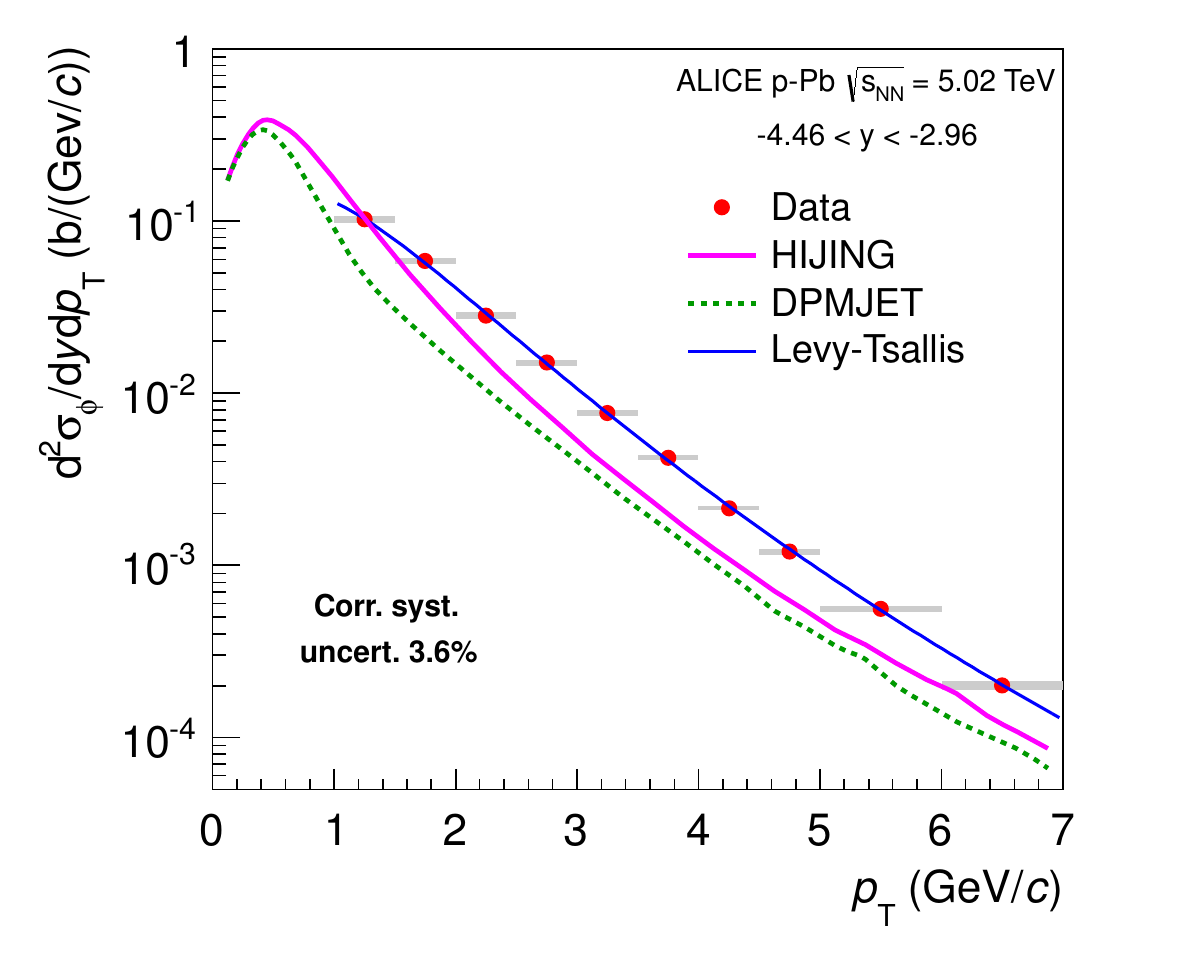}
    \includegraphics[width=0.49\textwidth]{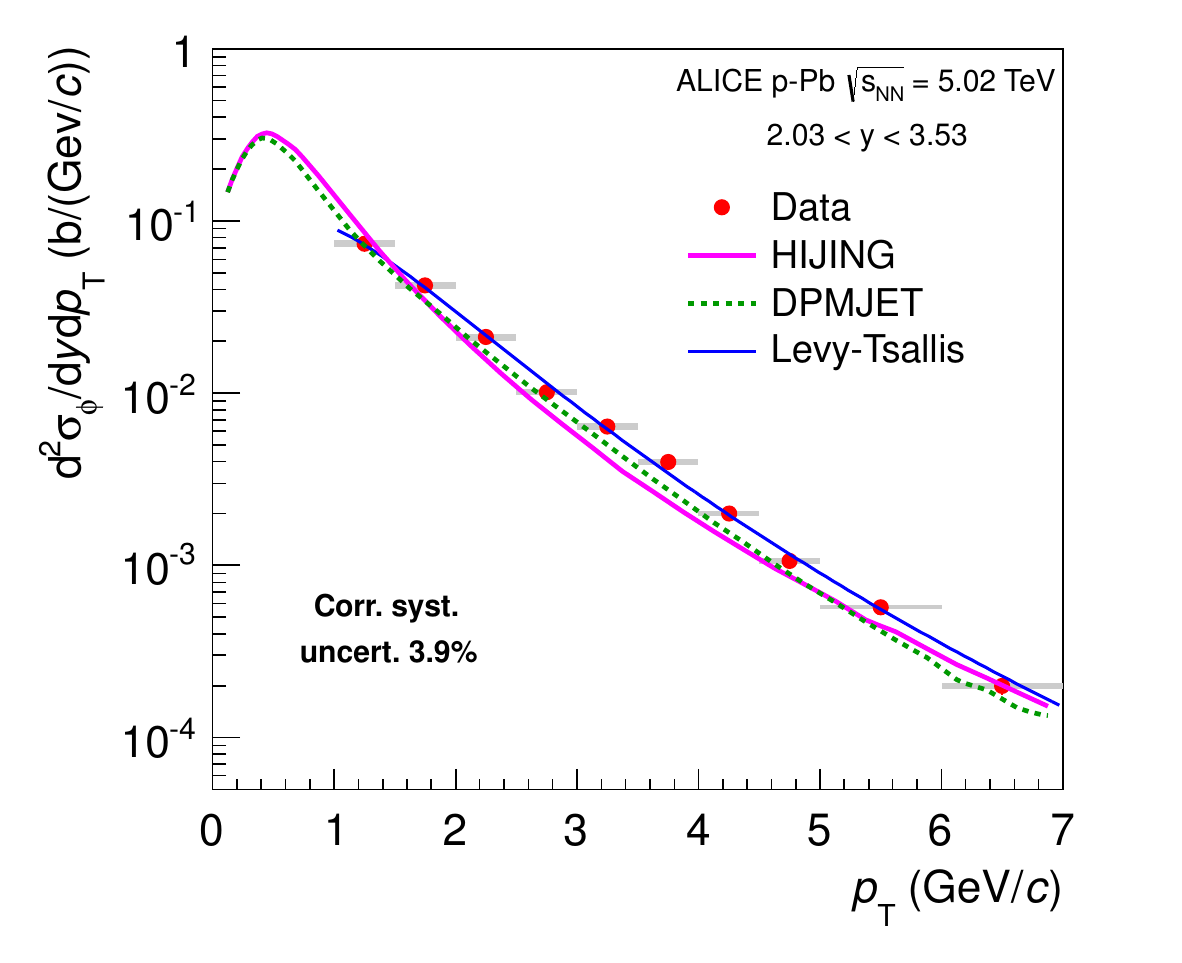} 
    \end{center} 
    \vspace{-0.5cm}
    \caption[\textwidth]{$\phi$-meson cross section in p-Pb collisions
      at $\snn = 5.02$~TeV as a function of $\pt$ in the backward (left) and forward (right)
    rapidity regions. Error bars (smaller
    than the markers) and boxes represent statistical and systematic uncertainties, respectively.
    Predictions by \texttt{HIJING}~\cite{Wang:1991hta} and \texttt{DPMJET}~\cite{Roesler:2000he} are also shown, together with 
    the result of a fit with the Levy-Tsallis function~(Eq.~\ref{eqn:ptLevy}).}
\label{fig:phiYieldVsPt}
\vspace{0.1cm}
\end{figure}

The $\phi$-meson cross section as a function of $\pt$ in p-Pb collisions is shown in \figurename~\ref{fig:phiYieldVsPt} 
for the forward and backward rapidity regions considered in the analysis.  
The results, also reported
in~\tablename~\ref{tab:pPbCrossSection_5.02}, are fitted with the 
Levy-Tsallis distribution defined in Eq.~(\ref{eqn:ptLevy}),
the resulting fit parameters being $\beta = 9.6 \pm 1.3$ and $T = 366 \pm 30$~MeV for the forward rapidity region and
$\beta = 11.4 \pm 1.4$ and $T = 384 \pm 24$~MeV
for the backward one, where the errors reflect the statistical uncertainties only. The predictions from
\texttt{HIJING} (with gluon shadowing)~\cite{Wang:1991hta} and \texttt{DPMJET}~\cite{Roesler:2000he} are
also shown: these generators provided a good description of the \ALICE
$\d N_\mathrm{ch}/\d\eta_\mathrm{lab}$ results at mid-rapidity~\cite{ALICE:2012xs}. 
Averaging over the available $\pt$~range, the discrepancy between the
data and the predictions from \texttt{HIJING} and \texttt{DPMJET}
amounts to $\sim 18\,\%$ and $\sim 57\,\%$, respectively, at backward
rapidity (the Pb-going direction) and
$\sim 5\,\%$ and $\sim 9.5\,\%$, respectively, at forward rapidity
(the p-going direction). In all the cases, the generators underestimate the data points.

\begin{table}[ht!]
  \begin{center}
    \begin{tabular}{l c c | c c}
    \hline \hline
    \raisebox{-8pt}[0pt][0pt]{$\pt$~(GeV/$c$)}~~~ & \multicolumn{2}{c |}{$-4.46 < y < -2.96$} & \multicolumn{2}{c}{$2.03 < y < 3.53$} \\ \cline{2-5} 
    ~& $\chi^2$/ndf & $\d^2 \sigma_\phi^\textrm{pPb} / (\d y \d \pt)$~(mb/(GeV/$c$)) & $\chi^2$/ndf & $\d^2 \sigma_\phi^\textrm{pPb} / (\d y \d \pt)$~(mb/(GeV/$c$)) \\ 
    \hline \hline
    $[1.0, 1.5]$    &  0.7 &  $102   \pm 8     \pm 12    $  &  1.5  &   $73.3   \pm 5.6   \pm 8.0   $  \\ 
    $[1.5, 2.0]$    &  1.2 &  $58.6  \pm 3.3   \pm 5.5   $  &  1.9  &   $42.1   \pm 2.5   \pm 4.3   $  \\ 
    $[2.0, 2.5]$    &  2.5 &  $28.3  \pm 1.4   \pm 2.9   $  &  1.7  &   $21.0   \pm 1.2   \pm 2.0   $  \\ 
    $[2.5, 3.0]$    &  4.2 &  $15.0  \pm 0.7   \pm 1.2   $  &  3.1  &   $10.07  \pm 0.77  \pm 0.97  $  \\ 
    $[3.0, 3.5]$    &  2.6 &  $7.66  \pm 0.40  \pm 0.70  $  &  2.0  &   $6.38   \pm 0.41  \pm 0.61  $  \\ 
    $[3.5, 4.0]$    &  1.9 &  $4.20  \pm 0.24  \pm 0.34  $  &  1.2  &   $3.96   \pm 0.30  \pm 0.36  $  \\
    $[4.0, 4.5]$    &  0.7 &  $2.15  \pm 0.17  \pm 0.16  $  &  1.0  &   $1.99   \pm 0.20  \pm 0.15  $  \\
    $[4.5, 5.0]$    &  0.9 &  $1.20  \pm 0.11  \pm 0.10  $  &  0.9  &   $1.06   \pm 0.13  \pm 0.08  $  \\
    $[5.0, 6.0]$    &  1.0 &  $0.560 \pm 0.052 \pm 0.054 $  &  1.0  &   $0.570  \pm 0.088 \pm 0.043 $  \\
    $[6.0, 7.0]$    &  1.2 &  $0.201 \pm 0.030 \pm 0.028 $  &  0.9  &   $0.199  \pm 0.045 \pm 0.016 $  \\         
    \hline \hline
    \end{tabular}
  \caption{Production cross section for the $\phi$-meson in p-Pb
    collisions at $\snn = 5.02$~TeV, as a function of $\pt$,
    in the backward and forward rapidity regions. The first uncertainty is statistical and the second is the bin-to-bin uncorrelated systematic. 
    The bin-to-bin correlated relative systematic uncertainty is 3.6\,\% and 3.9\,\%, respectively, for the backward and the forward regions.
    The $\chi^2$/ndf values are relative to the hadronic-cocktail fit and the [0.8,~1.2~GeV/$c^2$] mass region.}
  \label{tab:pPbCrossSection_5.02}
  \end{center}
\end{table}

\begin{figure}[t!] 
  \begin{center}
    \includegraphics[width=0.60\textwidth]{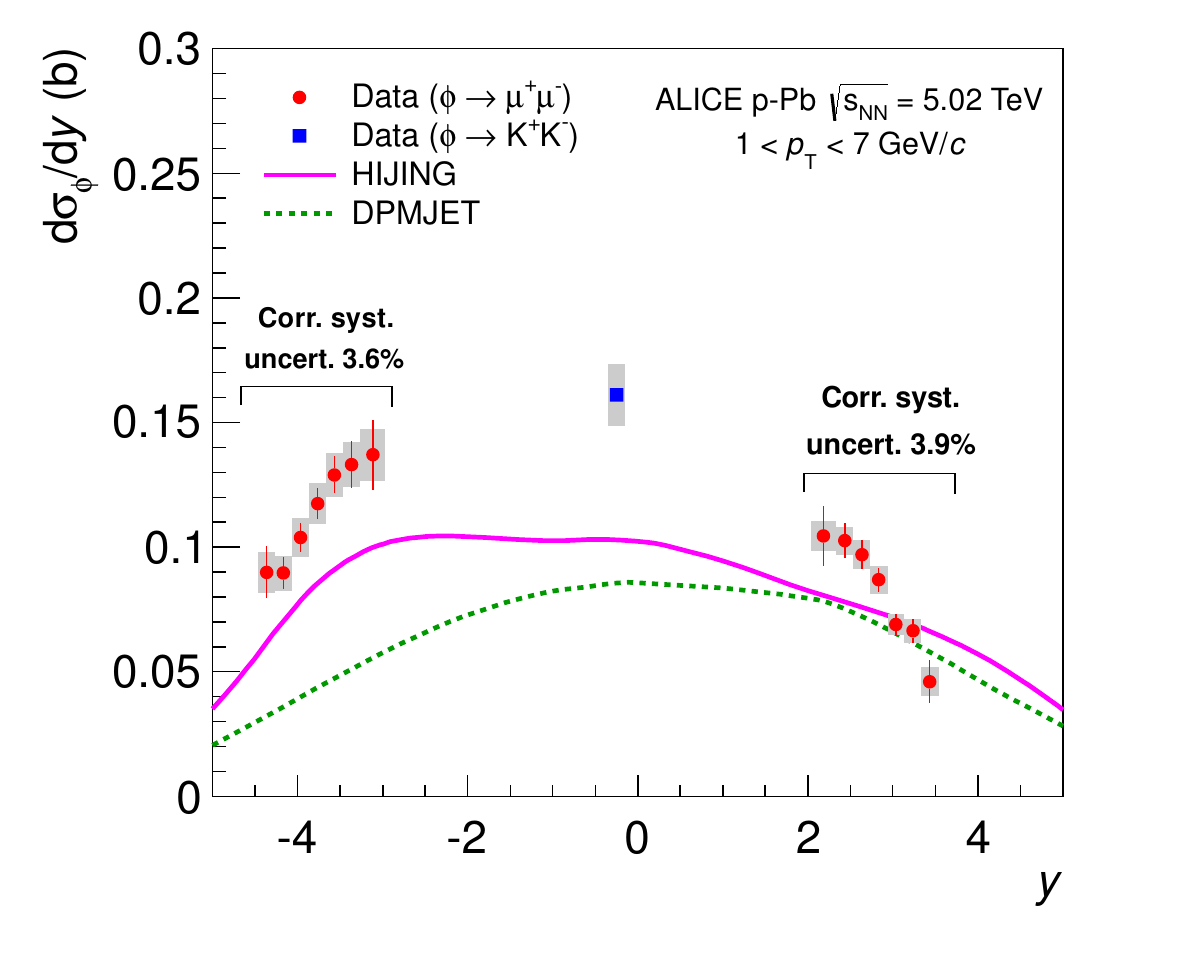}
  \end{center} 
  \vspace{-0.4cm}
  \caption[\textwidth]{$\phi$-meson cross section in p-Pb collisions 
    at $\snn = 5.02$~TeV as a function of rapidity, integrated over the range $1 < \pt < 7$~GeV/$c$.
    Error bars and boxes represent statistical and systematic uncertainties, respectively. 
    Predictions by \texttt{HIJING} and \texttt{DPMJET} are also shown, together with the mid-rapidity data point from the $\phi$-meson
    measurement in the K$^+$K$^-$ channel~\cite{Adam:2016bpr}, also evaluated in the range $1 < \pt < 7$~GeV/$c$.}
  \label{fig:phiYieldVsRapidity}
  \vspace{0.1cm}
\end{figure}

The $\phi$-meson cross section in p-Pb collisions, integrated over the accessible $\pt$ range, $1 < \pt < 7$~GeV/$c$, 
is shown as a function of rapidity in \figurename~\ref{fig:phiYieldVsRapidity}. 
The data points, also summarised in~\tablename~\ref{tab:pPbCrossSection_5.02_vs_y}, exhibit a significant asymmetry between 
the forward and backward rapidity regions. The data point
from the $\phi$-meson analysis at mid-rapidity in the K$^+$K$^-$ channel~\cite{Adam:2016bpr}, also shown for the $1 < \pt < 7$~GeV/$c$ $\pt$ range,
fits well into the trend defined by the two series of points in the backward and forward rapidity regions.
This observation complements the previous measurements of light-flavour particle production 
(charged unidentified particles) reported in p-Pb by \ALICE at the LHC at mid-rapidity~\cite{ALICE:2012xs}, 
and in d-Au by PHOBOS at RHIC ranging from mid to forward rapidity~\cite{Back:2003hx}. 
The comparison between the
data and the predictions by \texttt{HIJING} and \texttt{DPMJET}, illustrated in \figurename~\ref{fig:phiYieldVsRapidity},
clearly shows how the models --- which successfully described charged particle 
production at mid-rapidity in the same collision
system~\cite{ALICE:2012xs} --- fail to properly reproduce the shape and the normalisation of
the observed rapidity dependence of the $\phi$-meson cross section.
Still, the \texttt{HIJING} prediction qualitatively reproduces the forward-backward
asymmetry observed in the data, as well as --- ignoring the normalisation ---
the shape of the $y$-dependence in the backward region.
\texttt{DPMJET}, on the contrary, fails to reproduce even
qualitatively the observed forward-backward asymmetry. 

\begin{table}[ht!]
  \begin{center}
    \begin{tabular}{c c c || c c c }
    \hline \hline
    $y$ & $\chi^2$/ndf &~~$\d \sigma_\phi^\textrm{pPb} / \d y$~(mb) ~~& $y$ & $\chi^2$/ndf &~~$\d \sigma_\phi^\textrm{pPb} / \d y$~(mb) ~~ \\
    \hline \hline
    $[-4.46, ~-4.25]~~$ & 0.9 & $  89 \pm 10 \pm  9  $  &  $[2.03, ~2.35]~~$ & 2.6 & $ 104 \pm 11 \pm 6   $  \\
    $[-4.25, ~-4.05]~~$ & 1.8 & $  89 \pm 6  \pm  7  $  &  $[2.35, ~2.55]~~$ & 1.5 & $ 102 \pm 7  \pm 5   $  \\
    $[-4.05, ~-3.85]~~$ & 0.9 & $ 103 \pm 5  \pm  8  $  &  $[2.55, ~2.75]~~$ & 2.0 & $  96 \pm 5  \pm 6   $  \\
    $[-3.85, ~-3.65]~~$ & 2.9 & $ 117 \pm 6  \pm  9  $  &  $[2.75, ~2.95]~~$ & 1.6 & $  86 \pm 4  \pm 5   $  \\
    $[-3.65, ~-3.45]~~$ & 1.2 & $ 128 \pm 7  \pm  9  $  &  $[2.95, ~3.15]~~$ & 2.3 & $  68 \pm 4  \pm 4   $  \\
    $[-3.45, ~-3.25]~~$ & 3.6 & $ 133 \pm 9  \pm  9  $  &  $[3.15, ~3.35]~~$ & 1.0 & $  66 \pm 5  \pm 5   $  \\
    $[-3.25, ~-2.96]~~$ & 1.2 & $ 136 \pm 14 \pm 11  $  &  $[3.35, ~3.53]~~$ & 1.2 & $  45 \pm 8  \pm 6   $  \\
    \hline \hline
    \end{tabular}
  \caption{Production cross section for the $\phi$-meson in p-Pb collisions at $\snn = 5.02$~TeV, 
  as a function of rapidity, integrated over the range $1 < \pt <
  7$~GeV/$c$. The first uncertainty is statistical and the second is the bin-to-bin uncorrelated systematic. 
  The bin-to-bin correlated relative systematic uncertainty is 3.6\,\% and 3.9\,\%, respectively, for the backward and the forward regions.
  The $\chi^2$/ndf values are relative to the hadronic-cocktail fit and the 
  [0.8,~1.2~GeV/$c^2$] mass region.}
  \label{tab:pPbCrossSection_5.02_vs_y}
  \end{center}
\end{table}

\subsection{Forward-backward ratio in p-Pb collisions}

\noindent To establish a more direct comparison of the cross section
in the p-going and Pb-going directions, $\sigma_\phi^\textrm{pPb}$
was extracted as a function of $\pt$ in the common $|y|$ range $ 2.96<| y| < 3.53$.
The $\pt$ interval $1.0 < \pt < 1.5$~GeV/$c$ was
discarded in this measurement because of the poor statistics available
in this limited rapidity range, resulting in an uncertainty larger than $50\,\%$.

\begin{figure}[t!] 
    \begin{center}
    \includegraphics[width=0.60\textwidth]{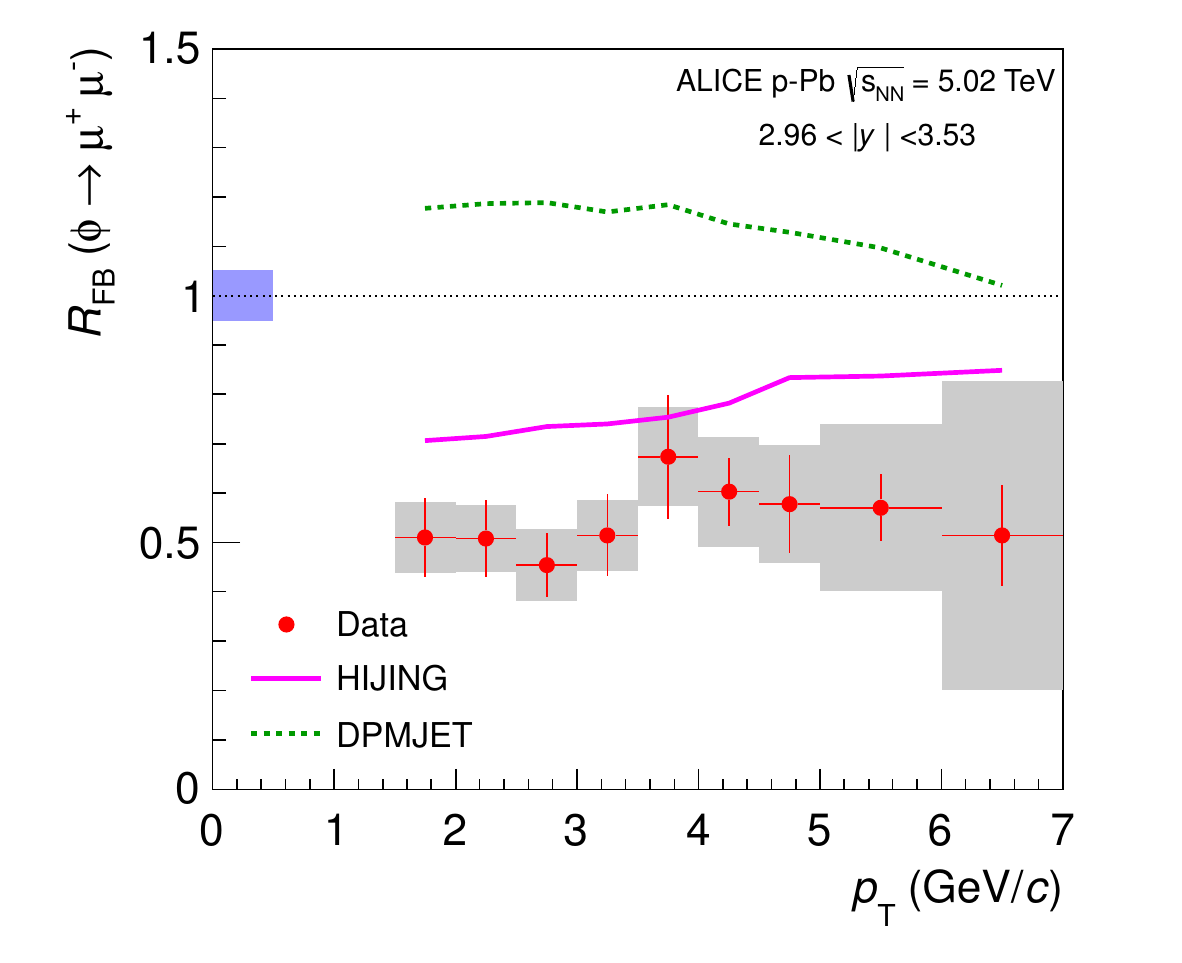}
    \end{center}
    \vspace{-0.4cm}
    \caption[\textwidth]{Forward-backward ratio for the $\phi$-meson
    in p-Pb collisions at $\snn = 5.02$~TeV as a function of $\pt$, 
    in the rapidity range $2.96 < |y| < 3.53$
    common to the two rapidity regions considered in the analysis. Error bars and boxes represent statistical and 
    systematic uncertainties, respectively. The blue box on the left represents the bin-to-bin
    correlated systematic uncertainty, see \tablename~\ref{tab:systematics}. Predictions from
    \texttt{HIJING} and \texttt{DPMJET} are also shown for comparison.}
\label{fig:forw_back_ratio_phi}
\vspace{0.1cm}
\end{figure}

The ratio between the forward and backward cross section, $\RFB$, is shown as a function of $\pt$ in \figurename~\ref{fig:forw_back_ratio_phi}.
The data points exhibit no significant $\pt$ dependence 
within the experimental uncertainties. Predictions by \texttt{HIJING}
and \texttt{DPMJET} are also shown, with \texttt{HIJING} slightly
overestimating the data points and \texttt{DPMJET} clearly failing to
reproduce the observed values, staying above $\RFB = 1$ in the whole
$\pt$ range considered here. This observation is consistent with the
observations in \figurename~\ref{fig:phiYieldVsRapidity}, where the
forward-backward asymmetry of the $\phi$-meson yield was better reproduced
by \texttt{HIJING} than by \texttt{DPMJET}.

\subsection{Nuclear modification factor in p-Pb collisions}

\noindent The $\phi$-meson nuclear modification factor $\RpPb$ is
defined as the ratio between the production cross section $\sigma_\phi^\mathrm{pPb}(\pt)$ in p-Pb collisions
and the cross section $\sigma_\phi^\mathrm{pp}(\pt)$ in pp collisions ---
evaluated at $\sqrt{s} = 5.02$~TeV as described in the following --- scaled by $A_\mathrm{Pb}$:
\begin{equation}
  \RpPb(\pt) = 
  \frac{\sigma_\phi^\mathrm{pPb}(\pt)}{\sigma_\phi^\mathrm{pp}(\pt) \cdot A_\mathrm{Pb}}~,
\end{equation}
where $A_\mathrm{Pb}$ is the nuclear mass number for the Pb nucleus.
Since for the pp cross section $\sigma_\phi^\mathrm{pp}$ at $\sqrt{s} = 5.02$~TeV
no direct measurement is currently available, it was evaluated by interpolating 
the measurements in the rapidity interval $2.5 < y < 4$ at
$\sqrt{s} = 2.76$ (see Section~\ref{cross_section_pp}) and 7~TeV~\cite{ALICE:2011ad}. 
For each $\pt$ interval, the $\sqrt{s}$ dependence of the differential cross section 
$\d^2\sigma_\phi^\mathrm{pp}/(\d y\, \d \pt)$ was described with a
power law 
$\sigma^\mathrm{pp}(\sqrt{s}) = C\cdot (\sqrt{s})^\alpha$, where $C$ and $\alpha$ are determined using the
data at 2.76 and 7~TeV. Alternative parameterisations were also considered~\cite{LHCb-CONF-2013-013}, namely 
a linear and an exponential function, and the mean of the results obtained with the three functions was taken.
Since the pp measurements are limited to $1 <\pt < 5$~GeV/$c$, the
cross section at $\snn = 5.02$~TeV was extrapolated towards higher $\pt$ by means of a Levy-Tsallis function, which describes the
calculated differential cross section in the $\pt$ range covered by the
measurements. The uncertainty on the interpolated cross sections arises from the choice of the function used for the interpolation,
from the uncertainties in the measurements at 2.76 and 7~TeV, and ---
for $\pt > 5$~GeV/$c$ --- from the extrapolation based on the
Levy-Tsallis fit. They range from about 7\,\%
for $\pt = 1$~GeV/$c$ to 20\,\% for $\pt = 5$~GeV/$c$, and exceed 30\,\% for $\pt > 5$~GeV/$c$, representing
the major source of systematic uncertainty for the measurement of the nuclear modification factor. 
The interpolated cross section, which refers to the rapidity range $2.5 < y < 4$,
was finally scaled to the forward and backward rapidity windows $2.03 < y < 3.53$
and $-4.46 < y < -2.96$, considered for the analysis of the p-Pb data. 
The relative scaling factors $f_\mathrm{fwd} = 1.135 \pm 0.031$ and 
$f_\mathrm{bkw} = 0.850 \pm 0.028$ were evaluated as an average from simulations with \texttt{PHOJET}
and the \texttt{Perugia0}, \texttt{Perugia11}, \texttt{ATLAS-CSC} and
\texttt{D6T} \texttt{PYTHIA} tunes. In doing so, we also retained the \texttt{PYTHIA} tunes
which were observed to fail in describing the pp data (see~Section~\ref{cross_section_pp}): the reason is
that the disagreement between models and data concerns in this case the absolute normalisation
more than the shape of the kinematic distributions, which is the only relevant feature 
in the evaluation of the $f_\mathrm{fwd}$ and $f_\mathrm{bkw}$ 
factors. The uncertainties (amounting to about 3\,\%) correspond to the differences 
between the considered MC~predictions. The numerical values are reported in
\tablename~\ref{tab:ppCrossSection_5.02}. 

\begin{table}[ht!]
  \begin{center}
    \begin{tabular}{l l l}
    \hline \hline
    \raisebox{-8pt}[0pt][0pt]{$\pt$~(GeV/$c$)}~~~~~~~~~ & \multicolumn{2}{c}{$\d^2 \sigma_\phi / \d y \d \pt$~(mb/(GeV/$c$))} \\ \cline{2-3}
    ~& $-4.46 < y < -2.96$~~~~~~~~~ & $2.03 < y < 3.53$ \\ 
    \hline \hline
    $[1.0, 1.5]$    &   $ 0.491     \pm 0.067 $       &     $ 0.656 \pm 0.090$ \\
    $[1.5, 2.0]$    &   $ 0.223     \pm 0.015 $       &     $ 0.297 \pm 0.020$ \\
    $[2.0, 2.5]$    &   $ 0.0995   \pm 0.0071 $     &     $ 0.1328 \pm 0.0095$ \\
    $[2.5, 3.0]$    &   $ 0.0467   \pm 0.0032 $     &     $ 0.0623 \pm 0.0043$ \\
    $[3.0, 3.5]$    &   $ 0.0234   \pm 0.0015 $     &     $ 0.0312 \pm 0.0020$ \\
    $[3.5, 4.0]$    &   $ 0.0125   \pm 0.0011 $     &     $ 0.0167 \pm 0.0015$ \\
    $[4.0, 4.5]$    &   $ 0.00706 \pm 0.00094 $   &     $ 0.0094 \pm 0.0012$ \\
    $[4.5, 5.0]$    &   $ 0.00419 \pm 0.00082 $   &     $ 0.0056 \pm 0.0011$ \\
    $[5.0, 6.0]$    &   $ 0.00213 \pm 0.00060 $   &     $ 0.00284 \pm 0.00081$ \\
    $[6.0, 7.0]$    &   $ 0.00093 \pm 0.00039 $   &     $ 0.00124 \pm 0.00052$ \\
    \hline \hline
    \end{tabular}
    \caption{Differential cross section for the $\phi$-meson in pp collisions at
      $\sqrt{s} = 5.02$~TeV in the backward and forward rapidity regions
      of interest for the analysis of the p-Pb data, as obtained
      interpolating the available measurements at $\sqrt{s} = 2.76$
      and 7~TeV. Total uncertainties, combining
      statistical and systematic sources, are reported.}
    \label{tab:ppCrossSection_5.02}
  \end{center}
\end{table}

\begin{figure}[t!] 
   \begin{center}
    \includegraphics[width=0.49\textwidth]{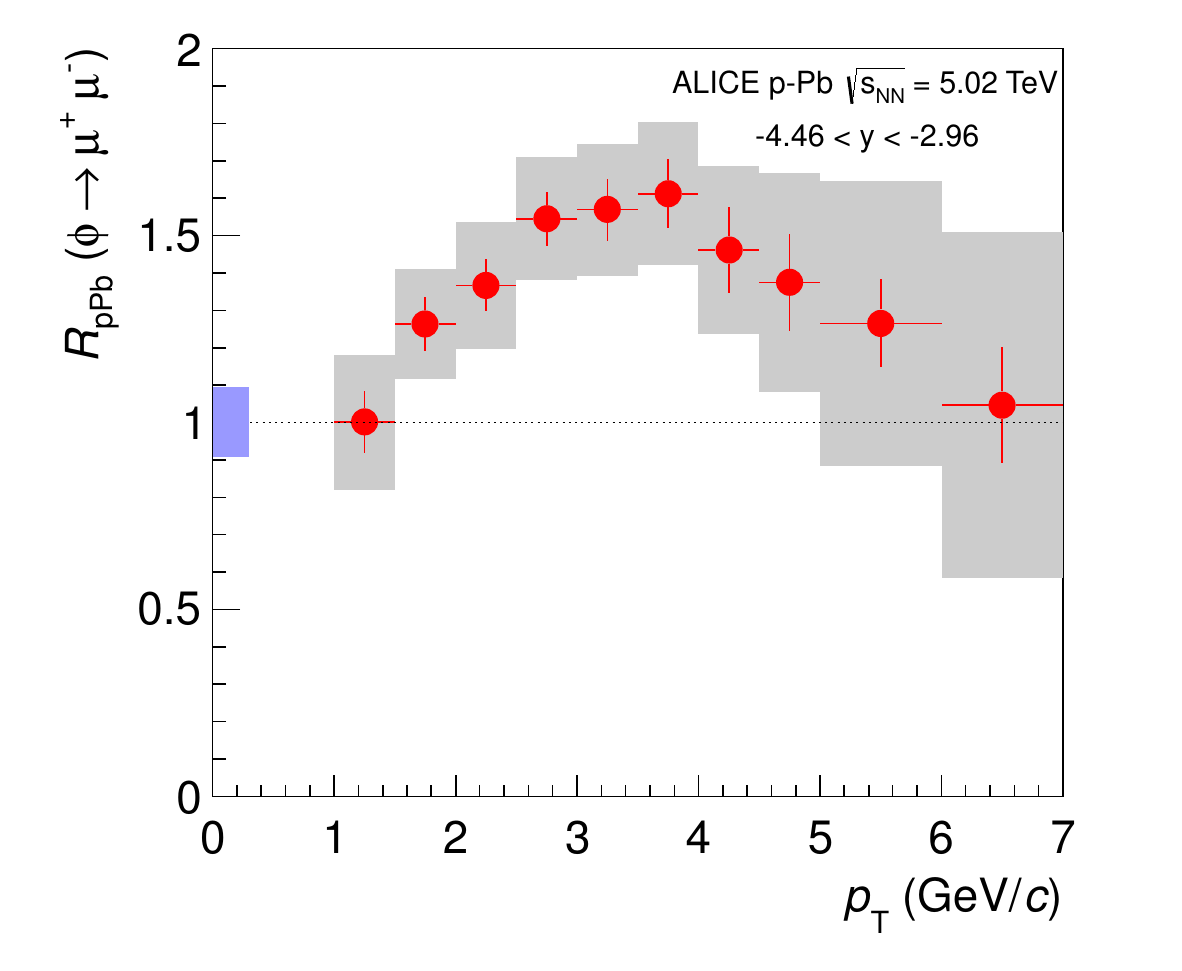}
    \includegraphics[width=0.49\textwidth]{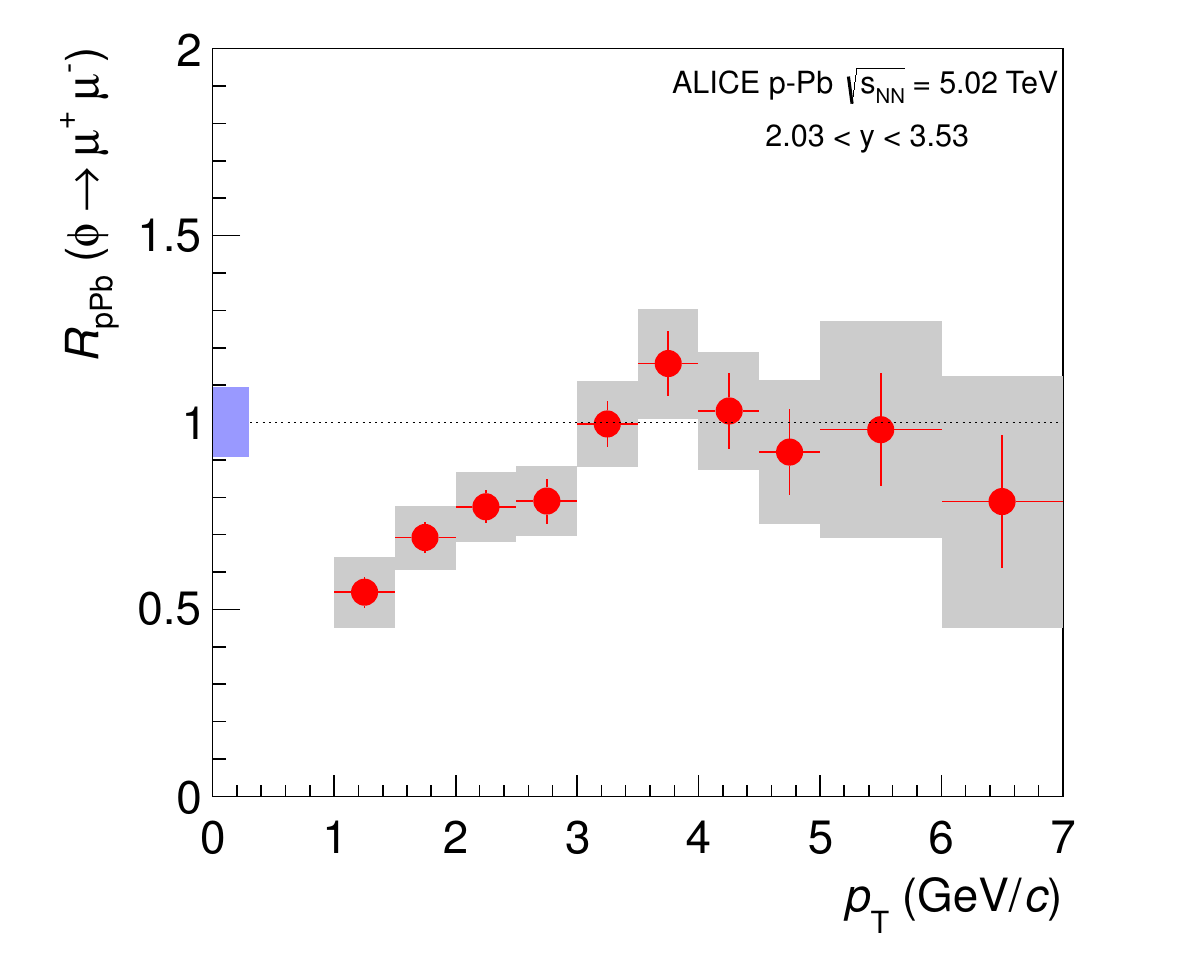} 
    \end{center} 
    \vspace{-0.2cm}
    \caption[\textwidth]{Nuclear modification factor $\RpPb$ in p-Pb
      collisions at $\snn = 5.02$~TeV 
    for the $\phi$-meson as a function of $\pt$, in the backward (left) and forward (right) 
    rapidity regions considered in the analysis. Error bars and boxes represent statistical and bin-to-bin uncorrelated systematic uncertainties, 
    respectively. The blue box on the left represents the bin-to-bin
    correlated systematic uncertainty, see \tablename~\ref{tab:systematics}.}
\label{fig:RpPb_vs_pt}
\vspace{0.5cm}
\end{figure}

The nuclear modification factor $\RpPb$ as a function of $\pt$ is shown in the two panels of \figurename~\ref{fig:RpPb_vs_pt}
for the backward and forward rapidity regions considered in the
analysis. The numerical values are also quoted
in \tablename~\ref{tab:R_pPb}. For each $\pt$ interval, the systematic uncertainty detailed
in \tablename~\ref{tab:systematics}
results from the quadratic sum of the uncertainty on the $\phi$-meson cross
section in p-Pb and the one of the 
pp~reference. A rising trend of $\RpPb$ when going from 
$\pt = 1$~GeV/$c$ to $\pt \approx 3$--4~GeV/$c$ can be observed both
at backward and forward rapidity. The values of $\RpPb$
in the two rapidity ranges, however, are significantly different. In particular, at backward rapidity 
we observe an enhancement of the $\phi$-meson cross section with respect to
the scaled pp reference peaked around
$\pt = 3$--4~GeV/$c$. This enhancement, absent in the forward rapidity
region, reaches a factor of up to
$\sim 1.6$ and could be associated either to an initial-state effect (including
a possible Cronin-like enhancement~\cite{Accardi:2002ik, Abt:2006wt}) or to a final state effect related to radial flow in p-Pb
as proposed for recent ALICE measurements at mid-rapidity~\cite{Adam:2016dau}.
Discriminating between these two effects requires more detailed investigations, 
including differential analyses as a function of global event properties like collision centrality.

Concerning the behaviour at high $\pt$, we observe that the $\phi$-meson 
$\RpPb$ is compatible with unity for $\pt \gtrsim 4$~GeV/$c$
in the p-going direction, similar to what was observed for the $\RpPb$ of charged particle production
at mid-rapidity~\cite{ALICE:2012mj,Adam:2016dau}. The observations in the Pb-going direction do not allow
a clear trend of the $\RpPb$ factor at high $\pt$ to be established.
A possible saturation at $\RpPb \approx 1$ for $\pt \gtrsim 5$~GeV/$c$ 
is, however, still compatible with the measurements.

\begin{table}[ht!]
  \begin{center}
    \begin{tabular}{l l l}
    \hline \hline
    \raisebox{-8pt}[0pt][0pt]{$\pt$~(GeV/$c$)}~~~~~~~~~ & \multicolumn{2}{c}{$\RpPb$} \\ \cline{2-3}
    ~& $-4.46 < y < -2.96$~~~~~~~~~ & $2.03 < y < 3.53$ \\ 
    \hline \hline
    $[1.0, 1.5]$    &     $1.00  \pm  0.08  \pm  0.18$     &     $0.537 \pm  0.041  \pm   0.094$  \\ 
    $[1.5, 2.0]$    &     $1.26  \pm  0.07  \pm  0.15$     &     $0.681 \pm  0.040  \pm   0.083$  \\ 
    $[2.0, 2.5]$    &     $1.37  \pm  0.07  \pm  0.17$     &     $0.760 \pm  0.043  \pm   0.091$  \\ 
    $[2.5, 3.0]$    &     $1.54  \pm  0.07  \pm  0.16$     &     $0.777 \pm  0.059  \pm   0.092$  \\ 
    $[3.0, 3.5]$    &     $1.57  \pm  0.08  \pm  0.18$     &     $0.98  \pm  0.06   \pm   0.11$   \\ 
    $[3.5, 4.0]$    &     $1.62  \pm  0.09  \pm  0.19$     &     $1.14  \pm  0.09   \pm   0.15$   \\
    $[4.0, 4.5]$    &     $1.46  \pm  0.12  \pm  0.22$     &     $1.02  \pm  0.10   \pm   0.15$   \\
    $[4.5, 5.0]$    &     $1.38  \pm  0.13  \pm  0.29$     &     $0.91  \pm  0.11   \pm   0.19$   \\
    $[5.0, 6.0]$    &     $1.26  \pm  0.12  \pm  0.38$     &     $0.97  \pm  0.15   \pm   0.29$   \\
    $[6.0, 7.0]$    &     $1.04  \pm  0.16  \pm  0.46$     &     $0.77  \pm  0.17   \pm   0.33$   \\     
    \hline \hline
    \end{tabular}
  \caption{Nuclear modification factor $\RpPb$ in p-Pb collisions at $\snn = 5.02$~TeV for the $\phi$-meson as a function of $\pt$ 
  in the backward and forward rapidity regions. The first uncertainty is statistical and the second is the bin-to-bin uncorrelated systematic.
  The bin-to-bin correlated relative systematic uncertainty is 8\,\%.}
  \label{tab:R_pPb}
  \end{center}
\end{table}

\begin{table}[ht!]
  \begin{center}
    \begin{tabular}{l c c}
    \hline \hline
    \raisebox{-8pt}[0pt][0pt]{Source}~~~~~~~~~~~~~~~~~~~~~~~~~~~ & \multicolumn{2}{c}{Syst.~uncertainty on $\sigma_\phi^\textrm{pPb}$~and~$\RpPb$} \\ \cline{2-3}
    ~& $-4.46 < y < -2.96$~~~~~~~~~ & $2.03 < y < 3.53$ \\ 
    \hline \hline
    \emph{Uncorrelated} & ~ & ~ \\
    \hline
    Signal extraction                &     3--5\,\%   &     4--8\,\%  \\ 
    Tracking efficiency              &     6\,\%     &       4\,\%  \\ 
    Trigger efficiency               &     3.2\,\%   &     2.8\,\%  \\ 
    $\sigma^\mathrm{pp}_\phi$        &     7--30\,\%  &    7--30\,\%  \\ 
    \hline \hline
   	\emph{Correlated} & ~ & ~ \\
    \hline
    $L_\mathrm{int}$                &     3.5\,\%  & 3.8\,\% \\
    $BR(\phi \to \ell\ell$)          &     1\,\%     &       1\,\%  \\ 
    $f_\mathrm{bkw}$ &     3.3\,\%  &      ---  \\ 
    $f_\mathrm{fwd}$ &     ---  &      2.7\,\%  \\ 
    \hline \hline
    \end{tabular}
  \caption{Systematic uncertainties (in percent) contributing to the measurement of the $\phi$-meson cross section 
  and nuclear modification factor in the backward and forward rapidity
  regions in p-Pb collisions at $\snn = 5.02$~TeV. When the uncertainty values depend on the $\pt$
  interval, their minimum and maximum values are quoted.}
  \label{tab:systematics}
  \end{center}
\end{table}

Only few other existing measurements can be compared to
our data. In particular, results on $\phi$-meson
production in d-Au collisions at $\snn = 200$~GeV have been recently released by the
PHENIX Collaboration~\cite{Adare:2015vvj}. The $\pt$-dependence of 
the $R_\mathrm{dAu}$ measured by PHENIX, as well as its evolution from
backward to forward rapidity, is found to be similar to what is observed
in our results for $\RpPb$. Mid-rapidity data on $R_\mathrm{dAu}$, also presented by
the PHENIX Collaboration for the $\phi$-meson, seem to sit between the forward- and
backward-rapidity results. Forward-rapidity measurements in d-Au collisions
at $\snn = 200$~GeV at RHIC~\cite{Arsene:2004ux, Back:2003hx} are also available for unidentified charged particles, 
although for the d-going direction only. These data exhibit, similar to our $\phi$-meson
results in the p-going direction, a rise of $R_\mathrm{dAu}$ from $\sim 0.5$ to $\sim 1$ 
between $\pt \sim 1$~GeV/$c$ and $\pt \sim 4$~GeV/$c$. 
A similar rise of $\RpPb$ in p-Pb collisions 
at $\snn = 5.02$~TeV is also observed in the already cited measurement of unidentified charged particle and identified 
charged pion and kaon production at mid-rapidity performed by \ALICE~\cite{ALICE:2012mj,Adam:2016dau}.
A recent study of $\phi$-meson production in p-Pb collisions at mid-rapidity by \ALICE~\cite{Adam:2016bpr} 
does not currently include results on~$\RpPb$.


\section{Conclusions}

\noindent We have presented results on $\phi$-meson production in the
dimuon channel in p-Pb collisions at $\snn = 5.02$~TeV obtained by the
ALICE experiment at the LHC.
Cross section and nuclear modification factor measurements were
performed for $1 < \pt < 7$~GeV/$c$ in the rapidity windows $2.03 < y
< 3.53$ (p-going direction) and $-4.46 < y < -2.96$ (Pb-going direction).
A corresponding cross section measurement in pp collisions at $\sqrt{s} = 2.76$~TeV 
has also been reported, for $1 < \pt < 5$~GeV/$c$ in the region $2.5 < y < 4$.
Predictions from \texttt{HIJING} and \texttt{DPMJET} are compared to
the p-Pb cross sections and are found to underestimate the data
both at backward (by about 18\,\% and 57\,\% on average, respectively) and at
forward rapidity (by about 5\,\% and 9.5\,\% on average, respectively).  
The forward-backward ratio in the $\phi$-meson cross section in p-Pb collisions was measured in the rapidity
range $2.96<|y|<3.53$, and no significant $\pt$ dependence was found within uncertainties.
In this case, the data points are significantly overestimated by the \texttt{DPMJET}
model, while only a slight disagreement is observed with respect
to the \texttt{HIJING} prediction.

In the p-going direction a
rising trend of the nuclear modification factor $\RpPb$ is observed
from $\sim 0.5$ to $\sim 1$, when going from $\pt = 1$~GeV/$c$ to $\pt
= 4$~GeV/$c$. This observation is
compatible with the behaviour of charged particles
at forward rapidity at RHIC energies, and at mid-rapidity at LHC energies. 
In the Pb-going direction, on the other hand, an enhancement is observed for $\RpPb$,
reaching values as large as $\sim 1.6$ around $\pt =
3$--4~GeV/$c$. An interpretation of these results, either in terms of an
initial-state (Cronin-like) effect or a final-state effect related
to radial flow in p-Pb, is not possible yet, due to a general lack of
theoretical predictions for particle
production in the light-flavour sector at forward rapidity in p-A collisions at the LHC energies.


\newenvironment{acknowledgement}{\relax}{\relax}

 \begin{acknowledgement}
 \section*{Acknowledgements}
The ALICE Collaboration would like to thank all its engineers and technicians for their invaluable contributions to the construction of the experiment and the CERN accelerator teams for the outstanding performance of the LHC complex.
The ALICE Collaboration gratefully acknowledges the resources and support provided by all Grid centres and the Worldwide LHC Computing Grid (WLCG) collaboration.
The ALICE Collaboration acknowledges the following funding agencies for their support in building and
running the ALICE detector:
State Committee of Science,  World Federation of Scientists (WFS)
and Swiss Fonds Kidagan, Armenia,
Conselho Nacional de Desenvolvimento Cient\'{\i}fico e Tecnol\'{o}gico (CNPq), Financiadora de Estudos e Projetos (FINEP),
Funda\c{c}\~{a}o de Amparo \`{a} Pesquisa do Estado de S\~{a}o Paulo (FAPESP);
National Natural Science Foundation of China (NSFC), the Chinese Ministry of Education (CMOE)
and the Ministry of Science and Technology of China (MSTC);
Ministry of Education and Youth of the Czech Republic;
Danish Natural Science Research Council, the Carlsberg Foundation and the Danish National Research Foundation;
The European Research Council under the European Community's Seventh Framework Programme;
Helsinki Institute of Physics and the Academy of Finland;
French CNRS-IN2P3, the `Region Pays de Loire', `Region Alsace', `Region Auvergne' and CEA, France;
German Bundesministerium fur Bildung, Wissenschaft, Forschung und Technologie (BMBF) and the Helmholtz Association;
General Secretariat for Research and Technology, Ministry of
Development, Greece;
Hungarian Orszagos Tudomanyos Kutatasi Alappgrammok (OTKA) and National Office for Research and Technology (NKTH);
Department of Atomic Energy and Department of Science and Technology of the Government of India;
Istituto Nazionale di Fisica Nucleare (INFN) and Centro Fermi -
Museo Storico della Fisica e Centro Studi e Ricerche "Enrico
Fermi", Italy;
MEXT Grant-in-Aid for Specially Promoted Research, Ja\-pan;
Joint Institute for Nuclear Research, Dubna;
National Research Foundation of Korea (NRF);
Consejo Nacional de Cienca y Tecnologia (CONACYT), Direccion General de Asuntos del Personal Academico(DGAPA), M\'{e}xico, :Amerique Latine Formation academique – European Commission(ALFA-EC) and the EPLANET Program
(European Particle Physics Latin American Network)
Stichting voor Fundamenteel Onderzoek der Materie (FOM) and the Nederlandse Organisatie voor Wetenschappelijk Onderzoek (NWO), Netherlands;
Research Council of Norway (NFR);
National Science Centre, Poland;
Ministry of National Education/Institute for Atomic Physics and Consiliul Naţional al Cercetării Ştiinţifice - Executive Agency for Higher Education Research Development and Innovation Funding (CNCS-UEFISCDI) - Romania;
Ministry of Education and Science of Russian Federation, Russian
Academy of Sciences, Russian Federal Agency of Atomic Energy,
Russian Federal Agency for Science and Innovations and The Russian
Foundation for Basic Research;
Ministry of Education of Slovakia;
Department of Science and Technology, South Africa;
Centro de Investigaciones Energeticas, Medioambientales y Tecnologicas (CIEMAT), E-Infrastructure shared between Europe and Latin America (EELA), Ministerio de Econom\'{i}a y Competitividad (MINECO) of Spain, Xunta de Galicia (Conseller\'{\i}a de Educaci\'{o}n),
Centro de Aplicaciones Tecnológicas y Desarrollo Nuclear (CEA\-DEN), Cubaenerg\'{\i}a, Cuba, and IAEA (International Atomic Energy Agency);
Swedish Research Council (VR) and Knut $\&$ Alice Wallenberg
Foundation (KAW);
Ukraine Ministry of Education and Science;
United Kingdom Science and Technology Facilities Council (STFC);
The United States Department of Energy, the United States National
Science Foundation, the State of Texas, and the State of Ohio;
Ministry of Science, Education and Sports of Croatia and  Unity through Knowledge Fund, Croatia.
Council of Scientific and Industrial Research (CSIR), New Delhi, India
 \end{acknowledgement}
\bibliographystyle{utphys}
\bibliography{ALICE_pA2013_phi}

\providecommand{\href}[2]{#2}\begingroup\raggedright\begin{thebibliography}{10}

\bibitem{Bazavov:2011nk}
A.~Bazavov, T.~Bhattacharya, M.~Cheng, C.~DeTar, H.~Ding, {\em et~al.}, ``{The
  chiral and deconfinement aspects of the QCD transition},''
  \href{http://dx.doi.org/10.1103/PhysRevD.85.054503}{{\em Phys.Rev.}
  {\bfseries D85} (2012) 054503},
\href{http://arxiv.org/abs/1111.1710}{{\ttfamily arXiv:1111.1710 [hep-lat]}}.

\bibitem{Borsanyi:2010bp}
{\bfseries Wuppertal-Budapest} Collaboration, S.~Borsanyi {\em et~al.}, ``{Is
  there still any $T_c$ mystery in lattice QCD? Results with physical masses in
  the continuum limit III},''
  \href{http://dx.doi.org/10.1007/JHEP09(2010)073}{{\em JHEP} {\bfseries 1009}
  (2010) 073},
\href{http://arxiv.org/abs/1005.3508}{{\ttfamily arXiv:1005.3508 [hep-lat]}}.

\bibitem{Borsanyi:2010cj}
S.~Borsanyi, G.~Endrodi, Z.~Fodor, A.~Jakovac, S.~D. Katz, {\em et~al.}, ``{The
  QCD equation of state with dynamical quarks},''
  \href{http://dx.doi.org/10.1007/JHEP11(2010)077}{{\em JHEP} {\bfseries 1011}
  (2010) 077},
\href{http://arxiv.org/abs/1007.2580}{{\ttfamily arXiv:1007.2580 [hep-lat]}}.

\bibitem{Accardi:2002ik}
A.~Accardi, ``{Cronin effect in proton nucleus collisions: A Survey of
  theoretical models},''
\href{http://arxiv.org/abs/hep-ph/0212148}{{\ttfamily arXiv:hep-ph/0212148
  [hep-ph]}}.

\bibitem{Salgado:2011wc}
C.~Salgado, J.~Alvarez-Muniz, F.~Arleo, N.~Armesto, M.~Botje, {\em et~al.},
  ``{Proton-Nucleus Collisions at the LHC: Scientific Opportunities and
  Requirements},'' \href{http://dx.doi.org/10.1088/0954-3899/39/1/015010}{{\em
  J.Phys.} {\bfseries G39} (2012) 015010},
\href{http://arxiv.org/abs/1105.3919}{{\ttfamily arXiv:1105.3919 [hep-ph]}}.

\bibitem{Brandt:2014vva}
M.~Brandt, M.~Klasen, and F.~König, ``{Nuclear parton density modifications
  from low-mass lepton pair production at the LHC},''
  \href{http://dx.doi.org/10.1016/j.nuclphysa.2014.03.024}{{\em Nucl. Phys.}
  {\bfseries A927} (2014) 78--90},
\href{http://arxiv.org/abs/1401.6817}{{\ttfamily arXiv:1401.6817 [hep-ph]}}.

\bibitem{Arsene:2004fa}
{\bfseries BRAHMS} Collaboration, I.~Arsene {\em et~al.}, ``{Quark gluon plasma
  and color glass condensate at RHIC? The Perspective from the BRAHMS
  experiment},'' \href{http://dx.doi.org/10.1016/j.nuclphysa.2005.02.130}{{\em
  Nucl.Phys.} {\bfseries A757} (2005) 1--27},
\href{http://arxiv.org/abs/nucl-ex/0410020}{{\ttfamily arXiv:nucl-ex/0410020
  [nucl-ex]}}.

\bibitem{Back:2004je}
B.~Back, M.~Baker, M.~Ballintijn, D.~Barton, B.~Becker, {\em et~al.}, ``{The
  PHOBOS perspective on discoveries at RHIC},''
  \href{http://dx.doi.org/10.1016/j.nuclphysa.2005.03.084}{{\em Nucl.Phys.}
  {\bfseries A757} (2005) 28--101},
\href{http://arxiv.org/abs/nucl-ex/0410022}{{\ttfamily arXiv:nucl-ex/0410022
  [nucl-ex]}}.

\bibitem{ALICE:2012xs}
{\bfseries ALICE} Collaboration, B.~Abelev {\em et~al.}, ``{Pseudorapidity
  density of charged particles $p$-Pb collisions at $\sqrt{s_{NN}}=5.02$
  TeV},'' \href{http://dx.doi.org/10.1103/PhysRevLett.110.032301}{{\em
  Phys.Rev.Lett.} {\bfseries 110} (2013) 032301},
\href{http://arxiv.org/abs/1210.3615}{{\ttfamily arXiv:1210.3615 [nucl-ex]}}.

\bibitem{ALICE:2012mj}
{\bfseries ALICE} Collaboration, B.~Abelev {\em et~al.}, ``{Transverse Momentum
  Distribution and Nuclear Modification Factor of Charged Particles in $p$-Pb
  Collisions at $\sqrt{s_{NN}}=5.02$ TeV},''
  \href{http://dx.doi.org/10.1103/PhysRevLett.110.082302}{{\em Phys.Rev.Lett.}
  {\bfseries 110} (2013) 082302},
\href{http://arxiv.org/abs/1210.4520}{{\ttfamily arXiv:1210.4520 [nucl-ex]}}.

\bibitem{Abelev:2014dsa}
{\bfseries ALICE} Collaboration, B.~Abelev {\em et~al.}, ``{Transverse momentum
  dependence of inclusive primary charged-particle production in p-Pb
  collisions at $\sqrt{s_\mathrm{{NN}}}=5.02~\text {TeV}$},''
  \href{http://dx.doi.org/10.1140/epjc/s10052-014-3054-5}{{\em Eur.Phys.J.}
  {\bfseries C74} no.~9, (2014) 3054},
\href{http://arxiv.org/abs/1405.2737}{{\ttfamily arXiv:1405.2737 [nucl-ex]}}.

\bibitem{Adam:2016dau}
{\bfseries ALICE} Collaboration, J.~Adam {\em et~al.}, ``{Multiplicity
  dependence of charged pion, kaon, and (anti)proton production at large
  transverse momentum in p-Pb collisions at $\mathbf{\sqrt{{\textit s}_{\rm
  NN}}}$ = 5.02 TeV},''
  \href{http://dx.doi.org/10.1016/j.physletb.2016.07.050}{{\em Phys. Lett.}
  {\bfseries B760} (2016) 720--735},
\href{http://arxiv.org/abs/1601.03658}{{\ttfamily arXiv:1601.03658 [nucl-ex]}}.

\bibitem{Arsene:2004ux}
{\bfseries BRAHMS} Collaboration, I.~Arsene {\em et~al.}, ``{On the evolution
  of the nuclear modification factors with rapidity and centrality in d + Au
  collisions at $\sqrt{s_{NN}}=200$~GeV},''
  \href{http://dx.doi.org/10.1103/PhysRevLett.93.242303}{{\em Phys.Rev.Lett.}
  {\bfseries 93} (2004) 242303},
\href{http://arxiv.org/abs/nucl-ex/0403005}{{\ttfamily arXiv:nucl-ex/0403005
  [nucl-ex]}}.

\bibitem{Back:2003hx}
{\bfseries PHOBOS} Collaboration, B.~Back {\em et~al.}, ``{Pseudorapidity
  distribution of charged particles in d + Au collisions at
  $\sqrt{s_{NN}}=200$~GeV},''
  \href{http://dx.doi.org/10.1103/PhysRevLett.93.082301}{{\em Phys.Rev.Lett.}
  {\bfseries 93} (2004) 082301},
\href{http://arxiv.org/abs/nucl-ex/0311009}{{\ttfamily arXiv:nucl-ex/0311009
  [nucl-ex]}}.

\bibitem{Rafelski:1982pu}
J.~Rafelski and B.~Muller, ``{Strangeness Production in the Quark - Gluon
  Plasma},''
\href{http://dx.doi.org/10.1103/PhysRevLett.48.1066}{{\em Phys.Rev.Lett.}
  {\bfseries 48} (1982) 1066}.

\bibitem{Shor:1984ui}
A.~Shor, ``{$\phi$-meson production as a probe of the Quark Gluon Plasma},''
\href{http://dx.doi.org/10.1103/PhysRevLett.54.1122}{{\em Phys.Rev.Lett.}
  {\bfseries 54} (1985) 1122--1125}.

\bibitem{Koch:1986ud}
P.~Koch, B.~Muller, and J.~Rafelski, ``{Strangeness in Relativistic Heavy Ion
  Collisions},''
\href{http://dx.doi.org/10.1016/0370-1573(86)90096-7}{{\em Phys.Rept.}
  {\bfseries 142} (1986) 167--262}.

\bibitem{Alt:2004wc}
{\bfseries NA49} Collaboration, C.~Alt {\em et~al.}, ``{System-size dependence
  of strangeness production in nucleus-nucleus collisions at s(NN)**(1/2) =
  17.3-GeV},'' \href{http://dx.doi.org/10.1103/PhysRevLett.94.052301}{{\em
  Phys. Rev. Lett.} {\bfseries 94} (2005) 052301},
\href{http://arxiv.org/abs/nucl-ex/0406031}{{\ttfamily arXiv:nucl-ex/0406031
  [nucl-ex]}}.

\bibitem{Alt:2008iv}
{\bfseries NA49} Collaboration, C.~Alt {\em et~al.}, ``{Energy dependence of
  phi meson production in central Pb+Pb collisions at s(NN)**(1/2) = 6 to 17
  GeV},'' \href{http://dx.doi.org/10.1103/PhysRevC.78.044907}{{\em Phys. Rev.}
  {\bfseries C78} (2008) 044907},
\href{http://arxiv.org/abs/0806.1937}{{\ttfamily arXiv:0806.1937 [nucl-ex]}}.

\bibitem{Alessandro:2003gy}
{\bfseries NA50} Collaboration, B.~Alessandro {\em et~al.}, ``{phi production
  in Pb - Pb collisions at 158-GeV/c per nucleon incident momentum},''
  \href{http://dx.doi.org/10.1016/S0370-2693(02)03267-7}{{\em Phys. Lett.}
  {\bfseries B555} (2003) 147--155}.
[Erratum: Phys. Lett.B561,294(2003)].

\bibitem{Banicz:2009aa}
{\bfseries NA60} Collaboration, K.~Banicz {\em et~al.}, ``{phi Production in
  In-In Collisions at 158-A-GeV},''
  \href{http://dx.doi.org/10.1140/epjc/s10052-009-1137-5}{{\em Eur. Phys. J.}
  {\bfseries C64} (2009) 1--18},
\href{http://arxiv.org/abs/0906.1102}{{\ttfamily arXiv:0906.1102 [hep-ex]}}.

\bibitem{Arnaldi:2011nn}
{\bfseries NA60} Collaboration, R.~Arnaldi {\em et~al.}, ``{A Comparative
  measurement of $\phi\rightarrow K^+K^-$ and $\phi\rightarrow \mu^+\mu^-$ in
  In-In collisions at the CERN SPS},''
  \href{http://dx.doi.org/10.1016/j.physletb.2011.04.028}{{\em Phys. Lett.}
  {\bfseries B699} (2011) 325--329},
\href{http://arxiv.org/abs/1104.4060}{{\ttfamily arXiv:1104.4060 [nucl-ex]}}.

\bibitem{Abelev:2008aa}
{\bfseries STAR} Collaboration, B.~Abelev {\em et~al.}, ``{Measurements of phi
  meson production in relativistic heavy-ion collisions at RHIC},''
  \href{http://dx.doi.org/10.1103/PhysRevC.79.064903}{{\em Phys. Rev.}
  {\bfseries C79} (2009) 064903},
\href{http://arxiv.org/abs/0809.4737}{{\ttfamily arXiv:0809.4737 [nucl-ex]}}.

\bibitem{Adler:2004hv}
{\bfseries PHENIX} Collaboration, S.~S. Adler {\em et~al.}, ``{Production of
  phi mesons at mid-rapidity in s(NN)**(1/2) = 200- GeV Au+Au collisions at
  RHIC},'' \href{http://dx.doi.org/10.1103/PhysRevC.72.014903}{{\em Phys. Rev.}
  {\bfseries C72} (2005) 014903},
\href{http://arxiv.org/abs/nucl-ex/0410012}{{\ttfamily arXiv:nucl-ex/0410012
  [nucl-ex]}}.

\bibitem{ALICE:2011ad}
{\bfseries ALICE} Collaboration, B.~Abelev {\em et~al.}, ``{Light vector meson
  production in $pp$ collisions at $\sqrt{s}=7$~TeV},''
  \href{http://dx.doi.org/10.1016/j.physletb.2012.03.038}{{\em Phys.Lett.}
  {\bfseries B710} (2012) 557--568},
\href{http://arxiv.org/abs/1112.2222}{{\ttfamily arXiv:1112.2222 [nucl-ex]}}.

\bibitem{Aamodt:2008zz}
{\bfseries ALICE} Collaboration, K.~Aamodt {\em et~al.}, ``{The ALICE
  experiment at the CERN LHC},''
\href{http://dx.doi.org/10.1088/1748-0221/3/08/S08002}{{\em JINST} {\bfseries
  3} (2008) S08002}.

\bibitem{Abelev:2014ffa}
{\bfseries ALICE} Collaboration, B.~Abelev {\em et~al.}, ``{Performance of the
  ALICE Experiment at the CERN LHC},''
  \href{http://dx.doi.org/10.1142/S0217751X14300440}{{\em Int.J.Mod.Phys.}
  {\bfseries A29} (2014) 1430044},
\href{http://arxiv.org/abs/1402.4476}{{\ttfamily arXiv:1402.4476 [nucl-ex]}}.

\bibitem{Gagliardi:2011he}
{\bfseries ALICE} Collaboration, M.~Gagliardi, ``{Measurement of reference
  cross sections in pp and Pb-Pb collisions at the LHC in van der Meer scans
  with the ALICE detector},'' \href{http://dx.doi.org/10.1063/1.3692205}{{\em
  AIP Conf.Proc.} {\bfseries 1422} (2012) 110--116},
\href{http://arxiv.org/abs/1109.5369}{{\ttfamily arXiv:1109.5369 [hep-ex]}}.

\bibitem{Abelev:2013yxa}
{\bfseries ALICE} Collaboration, B.~Abelev {\em et~al.}, ``{$J/\psi$ production
  and nuclear effects in p-Pb collisions at $\sqrt{s_{NN}}$ = 5.02 TeV},''
  \href{http://dx.doi.org/10.1007/JHEP02(2014)073}{{\em JHEP} {\bfseries 1402}
  (2014) 073},
\href{http://arxiv.org/abs/1308.6726}{{\ttfamily arXiv:1308.6726 [nucl-ex]}}.

\bibitem{Abelev:2014epa}
{\bfseries ALICE} Collaboration, B.~Abelev {\em et~al.}, ``{Measurement of
  visible cross sections in proton-lead collisions at $\sqrt{s_{\rm NN}}$ =
  5.02 TeV in van der Meer scans with the ALICE detector},''
  \href{http://dx.doi.org/10.1088/1748-0221/9/11/P11003}{{\em JINST} {\bfseries
  9} no.~11, (2014) P11003},
\href{http://arxiv.org/abs/1405.1849}{{\ttfamily arXiv:1405.1849 [nucl-ex]}}.

\bibitem{Abelev:2014qha}
{\bfseries ALICE} Collaboration, B.~Abelev {\em et~al.}, ``{Measurement of
  quarkonium production at forward rapidity in $pp$ collisions at $\sqrt{s} =
  7$ TeV},'' \href{http://dx.doi.org/10.1140/epjc/s10052-014-2974-4}{{\em
  Eur.Phys.J.} {\bfseries C74} no.~8, (2014) 2974},
\href{http://arxiv.org/abs/1403.3648}{{\ttfamily arXiv:1403.3648 [nucl-ex]}}.

\bibitem{Abelev:2012sea}
{\bfseries ALICE} Collaboration, B.~Abelev {\em et~al.}, ``{Measurement of
  inelastic, single- and double-diffraction cross sections in proton--proton
  collisions at the LHC with ALICE},''
  \href{http://dx.doi.org/10.1140/epjc/s10052-013-2456-0}{{\em Eur.Phys.J.}
  {\bfseries C73} no.~6, (2013) 2456},
\href{http://arxiv.org/abs/1208.4968}{{\ttfamily arXiv:1208.4968 [hep-ex]}}.

\bibitem{OfflineNote}
{\bfseries ALICE} Collaboration, L.~Aphecetche {\em et~al.}, ``{Numerical
  Simulations and Offline Reconstruction of the Muon Spectrometer of ALICE,
  ALICE-INT-2009-044},''.

\bibitem{kalmanC}
{\bfseries ALICE} Collaboration, G.~Chabratova {\em et~al.}, ``{Development of
  the Kalman filter for tracking in the forward muon spectrometer of ALICE,
  ALICE-INT-2003-002},''.

\bibitem{ALICE-PUBLIC-2016-004}
{\bfseries ALICE} Collaboration, ``{Supplemental figures for ``$\phi$-meson
  production at forward rapidity in p-Pb collisions at $\sqrt{s_{\rm NN}}=
  5.02$ TeV and in pp collisions at $\sqrt{s}= 2.76$ TeV''},''.
  \url{https://cds.cern.ch/record/2195308}.

\bibitem{Sjostrand:2006za}
T.~Sjostrand, S.~Mrenna, and P.~Z. Skands, ``{PYTHIA 6.4 Physics and Manual},''
  \href{http://dx.doi.org/10.1088/1126-6708/2006/05/026}{{\em JHEP} {\bfseries
  0605} (2006) 026},
\href{http://arxiv.org/abs/hep-ph/0603175}{{\ttfamily arXiv:hep-ph/0603175
  [hep-ph]}}.

\bibitem{Brun:1994aa}
R.~Brun, F.~Carminati, and S.~Giani,
``{GEANT Detector Description and Simulation Tool, CERN-W-5013, 1994},''.

\bibitem{Beringer:1900zz}
{\bfseries Particle Data Group} Collaboration, J.~Beringer {\em et~al.},
  ``{Review of Particle Physics (RPP)},''
\href{http://dx.doi.org/10.1103/PhysRevD.86.010001}{{\em Phys.Rev.} {\bfseries
  D86} (2012) 010001}.

\bibitem{Oreglia:1980cs}
M.~Oreglia, {\em {A Study of the Reactions $\psi^\prime \to \gamma \gamma
  \psi$}}.
\newblock PhD thesis, SLAC, 1980.
\newblock
\url{http://search.proquest.com/docview/303036283}.
\newblock

\bibitem{Aamodt:2011gj}
{\bfseries ALICE} Collaboration, K.~Aamodt {\em et~al.}, ``{Rapidity and
  transverse momentum dependence of inclusive J$/\psi$ production in $pp$
  collisions at $\sqrt{s} = 7$ TeV},''
  \href{http://dx.doi.org/10.1016/j.physletb.2011.09.054,
  10.1016/j.physletb.2012.10.060}{{\em Phys. Lett.} {\bfseries B704} (2011)
  442--455}, \href{http://arxiv.org/abs/1105.0380}{{\ttfamily arXiv:1105.0380
  [hep-ex]}}.
[Erratum: Phys. Lett.B718,692(2012)].

\bibitem{Barlow:2002yb}
R.~Barlow, ``{Systematic errors: Facts and fictions},''
\href{http://arxiv.org/abs/hep-ex/0207026}{{\ttfamily arXiv:hep-ex/0207026
  [hep-ex]}}.

\bibitem{Engel:1995yda}
R.~Engel and J.~Ranft, ``{Hadronic photon-photon interactions at
  high-energies},'' \href{http://dx.doi.org/10.1103/PhysRevD.54.4244}{{\em
  Phys.Rev.} {\bfseries D54} (1996) 4244--4262},
\href{http://arxiv.org/abs/hep-ph/9509373}{{\ttfamily arXiv:hep-ph/9509373
  [hep-ph]}}.

\bibitem{Skands:2010ak}
P.~Z. Skands, ``{Tuning Monte Carlo Generators: The Perugia Tunes},''
  \href{http://dx.doi.org/10.1103/PhysRevD.82.074018}{{\em Phys.Rev.}
  {\bfseries D82} (2010) 074018},
\href{http://arxiv.org/abs/1005.3457}{{\ttfamily arXiv:1005.3457 [hep-ph]}}.

\bibitem{Buttar:2004iy}
C.~Buttar, D.~Clements, I.~Dawson, and A.~Moraes, ``{Simulations of minimum
  bias events and the underlying event, MC tuning and predictions for the
  LHC},''
{\em Acta Phys.Polon.} {\bfseries B35} (2004) 433--441.

\bibitem{Field:2008zz}
R.~Field, ``{Physics at the Tevatron},''
{\em Acta Phys.Polon.} {\bfseries B39} (2008) 2611--2672.

\bibitem{Abelev:2012hy}
{\bfseries ALICE} Collaboration, B.~Abelev {\em et~al.}, ``{Production of
  $K^*(892)^0$ and $\phi(1020)$ in $pp$ collisions at $\sqrt{s}=7$ TeV},''
  \href{http://dx.doi.org/10.1140/epjc/s10052-012-2183-y}{{\em Eur.Phys.J.}
  {\bfseries C72} (2012) 2183},
\href{http://arxiv.org/abs/1208.5717}{{\ttfamily arXiv:1208.5717 [hep-ex]}}.

\bibitem{Chatrchyan:2012qb}
{\bfseries CMS} Collaboration, S.~Chatrchyan {\em et~al.}, ``{Study of the
  inclusive production of charged pions, kaons, and protons in $pp$ collisions
  at $\sqrt{s}=0.9$, 2.76, and 7 TeV},''
  \href{http://dx.doi.org/10.1140/epjc/s10052-012-2164-1}{{\em Eur.Phys.J.}
  {\bfseries C72} (2012) 2164},
\href{http://arxiv.org/abs/1207.4724}{{\ttfamily arXiv:1207.4724 [hep-ex]}}.

\bibitem{Tsallis:1987eu}
C.~Tsallis, ``{Possible Generalization of Boltzmann-Gibbs Statistics},''
\href{http://dx.doi.org/10.1007/BF01016429}{{\em J.Statist.Phys.} {\bfseries
  52} (1988) 479--487}.

\bibitem{Wang:1991hta}
X.-N. Wang and M.~Gyulassy, ``{HIJING: A Monte Carlo model for multiple jet
  production in p p, p A and A A collisions},''
\href{http://dx.doi.org/10.1103/PhysRevD.44.3501}{{\em Phys.Rev.} {\bfseries
  D44} (1991) 3501--3516}.

\bibitem{Roesler:2000he}
S.~Roesler, R.~Engel, and J.~Ranft,
  \href{http://dx.doi.org/10.1007/978-3-642-18211-2_166}{``{The Monte Carlo
  event generator DPMJET-III},''} in {\em {Advanced Monte Carlo for radiation
  physics, particle transport simulation and applications. Proceedings,
  Conference, MC2000, Lisbon, Portugal, October 23-26, 2000}}, pp.~1033--1038.
\newblock 2000.
\newblock \href{http://arxiv.org/abs/hep-ph/0012252}{{\ttfamily
  arXiv:hep-ph/0012252 [hep-ph]}}.
\newblock
\url{http://www-public.slac.stanford.edu/sciDoc/docMeta.aspx?slacPubNumber=SLAC-PUB-8740}.
\newblock

\bibitem{Adam:2016bpr}
{\bfseries ALICE} Collaboration, J.~Adam {\em et~al.}, ``{Production of K$^{*}$
  (892)$^{0}$ and $\phi $ (1020) in p?Pb collisions at $\sqrt{s_{{\text
  {NN}}}}$ = 5.02 TeV},''
  \href{http://dx.doi.org/10.1140/epjc/s10052-016-4088-7}{{\em Eur. Phys. J.}
  {\bfseries C76} no.~5, (2016) 245},
\href{http://arxiv.org/abs/1601.07868}{{\ttfamily arXiv:1601.07868 [nucl-ex]}}.

\bibitem{LHCb-CONF-2013-013}
{\bfseries ALICE, LHCb} Collaboration, ``{Reference $pp$ cross-sections for
  $J/\psi$ studies in proton-lead collisions at $\sqrt{s_{NN}} = 5.02$~TeV and
  comparisons between ALICE and LHCb results},''.
  \url{http://cds.cern.ch/record/1639617}. CONF-2013-013.

\bibitem{Abt:2006wt}
{\bfseries HERA-B} Collaboration, I.~Abt {\em et~al.}, ``{K*0 and phi meson
  production in proton-nucleus interactions at s**(1/2) = 41.6-GeV},''
  \href{http://dx.doi.org/10.1140/epjc/s10052-007-0237-3}{{\em Eur.Phys.J.}
  {\bfseries C50} (2007) 315--328},
\href{http://arxiv.org/abs/hep-ex/0606049}{{\ttfamily arXiv:hep-ex/0606049
  [hep-ex]}}.

\bibitem{Adare:2015vvj}
{\bfseries PHENIX} Collaboration, A.~Adare {\em et~al.}, ``{$\phi$ meson
  production in $d$$+$Au collisions at $\sqrt{s_{_{NN}}}=200$ GeV},''
  \href{http://dx.doi.org/10.1103/PhysRevC.92.044909}{{\em Phys. Rev.}
  {\bfseries C92} no.~4, (2015) 044909},
\href{http://arxiv.org/abs/1506.08181}{{\ttfamily arXiv:1506.08181 [nucl-ex]}}.

\end{thebibliography}\endgroup
\newpage
%
%
\appendix
\section{The ALICE Collaboration}
\label{app:collab}



\begingroup
\small
\begin{flushleft}
J.~Adam\Irefn{org40}\And
D.~Adamov\'{a}\Irefn{org83}\And
M.M.~Aggarwal\Irefn{org87}\And
G.~Aglieri Rinella\Irefn{org36}\And
M.~Agnello\Irefn{org111}\And
N.~Agrawal\Irefn{org48}\And
Z.~Ahammed\Irefn{org132}\And
S.U.~Ahn\Irefn{org68}\And
I.~Aimo\Irefn{org94}\textsuperscript{,}\Irefn{org111}\And
S.~Aiola\Irefn{org136}
% ...
\renewcommand\labelenumi{\textsuperscript{\theenumi}~}

\section*{Affiliation notes}
\renewcommand\theenumi{\roman{enumi}}
\begin{Authlist}
\item \Adef{0}Deceased
\item \Adef{idp3815824}{Also at: M.V. Lomonosov Moscow State University, D.V. Skobeltsyn Institute of Nuclear, Physics, Moscow, Russia}
\item \Adef{idp5963136}{Also at: University of Kansas, Lawrence, Kansas, United States}
\end{Authlist}

\section*{Collaboration Institutes}
\renewcommand\theenumi{\arabic{enumi}~}
\begin{Authlist}

\item \Idef{org1}A.I. Alikhanyan National Science Laboratory (Yerevan Physics Institute) Foundation, Yerevan, Armenia
\item \Idef{org2}Benem\'{e}rita Universidad Aut\'{o}noma de Puebla, Puebla, Mexico
\item \Idef{org3}Bogolyubov Institute for Theoretical Physics, Kiev, Ukraine
\item \Idef{org4}Bose Institute, Department of Physics and Centre for Astroparticle Physics and Space Science (CAPSS), Kolkata, India
\item \Idef{org5}Budker Institute for Nuclear Physics, Novosibirsk, Russia
\item \Idef{org6}California Polytechnic State University, San Luis Obispo, California, United States
\item \Idef{org7}Central China Normal University, Wuhan, China
\item \Idef{org8}Centre de Calcul de l'IN2P3, Villeurbanne, France
\item \Idef{org9}Centro de Aplicaciones Tecnol\'{o}gicas y Desarrollo Nuclear (CEADEN), Havana, Cuba
\item \Idef{org10}Centro de Investigaciones Energ\'{e}ticas Medioambientales y Tecnol\'{o}gicas (CIEMAT), Madrid, Spain
\item \Idef{org11}Centro de Investigaci\'{o}n y de Estudios Avanzados (CINVESTAV), Mexico City and M\'{e}rida, Mexico
\item \Idef{org12}Centro Fermi - Museo Storico della Fisica e Centro Studi e Ricerche ``Enrico Fermi'', Rome, Italy
\item \Idef{org13}Chicago State University, Chicago, Illinois, USA
\item \Idef{org14}China Institute of Atomic Energy, Beijing, China
\item \Idef{org15}Commissariat \`{a} l'Energie Atomique, IRFU, Saclay, France
\item \Idef{org16}COMSATS Institute of Information Technology (CIIT), Islamabad, Pakistan
\item \Idef{org17}Departamento de F\'{\i}sica de Part\'{\i}culas and IGFAE, Universidad de Santiago de Compostela, Santiago de Compostela, Spain
\item \Idef{org18}Department of Physics and Technology, University of Bergen, Bergen, Norway
\item \Idef{org19}Department of Physics, Aligarh Muslim University, Aligarh, India
\item \Idef{org20}Department of Physics, Ohio State University, Columbus, Ohio, United States
\item \Idef{org21}Department of Physics, Sejong University, Seoul, South Korea
\item \Idef{org22}Department of Physics, University of Oslo, Oslo, Norway
\item \Idef{org23}Dipartimento di Elettrotecnica ed Elettronica del Politecnico, Bari, Italy
\item \Idef{org24}Dipartimento di Fisica dell'Universit\`{a} 'La Sapienza' and Sezione INFN Rome, Italy
\item \Idef{org25}Dipartimento di Fisica dell'Universit\`{a} and Sezione INFN, Cagliari, Italy
\item \Idef{org26}Dipartimento di Fisica dell'Universit\`{a} and Sezione INFN, Trieste, Italy
\item \Idef{org27}Dipartimento di Fisica dell'Universit\`{a} and Sezione INFN, Turin, Italy
\item \Idef{org28}Dipartimento di Fisica e Astronomia dell'Universit\`{a} and Sezione INFN, Bologna, Italy
\item \Idef{org29}Dipartimento di Fisica e Astronomia dell'Universit\`{a} and Sezione INFN, Catania, Italy
\item \Idef{org30}Dipartimento di Fisica e Astronomia dell'Universit\`{a} and Sezione INFN, Padova, Italy
\item \Idef{org31}Dipartimento di Fisica `E.R.~Caianiello' dell'Universit\`{a} and Gruppo Collegato INFN, Salerno, Italy
\item \Idef{org32}Dipartimento di Scienze e Innovazione Tecnologica dell'Universit\`{a} del  Piemonte Orientale and Gruppo Collegato INFN, Alessandria, Italy
\item \Idef{org33}Dipartimento Interateneo di Fisica `M.~Merlin' and Sezione INFN, Bari, Italy
\item \Idef{org34}Division of Experimental High Energy Physics, University of Lund, Lund, Sweden
\item \Idef{org35}Eberhard Karls Universit\"{a}t T\"{u}bingen, T\"{u}bingen, Germany
\item \Idef{org36}European Organization for Nuclear Research (CERN), Geneva, Switzerland
\item \Idef{org37}Excellence Cluster Universe, Technische Universit\"{a}t M\"{u}nchen, Munich, Germany
\item \Idef{org38}Faculty of Engineering, Bergen University College, Bergen, Norway
\item \Idef{org39}Faculty of Mathematics, Physics and Informatics, Comenius University, Bratislava, Slovakia
\item \Idef{org40}Faculty of Nuclear Sciences and Physical Engineering, Czech Technical University in Prague, Prague, Czech Republic
\item \Idef{org41}Faculty of Science, P.J.~\v{S}af\'{a}rik University, Ko\v{s}ice, Slovakia
\item \Idef{org42}Faculty of Technology, Buskerud and Vestfold University College, Vestfold, Norway
\item \Idef{org43}Frankfurt Institute for Advanced Studies, Johann Wolfgang Goethe-Universit\"{a}t Frankfurt, Frankfurt, Germany
\item \Idef{org44}Gangneung-Wonju National University, Gangneung, South Korea
\item \Idef{org45}Gauhati University, Department of Physics, Guwahati, India
\item \Idef{org46}Helsinki Institute of Physics (HIP), Helsinki, Finland
\item \Idef{org47}Hiroshima University, Hiroshima, Japan
\item \Idef{org48}Indian Institute of Technology Bombay (IIT), Mumbai, India
\item \Idef{org49}Indian Institute of Technology Indore, Indore (IITI), India
\item \Idef{org50}Inha University, Incheon, South Korea
\item \Idef{org51}Institut de Physique Nucl\'eaire d'Orsay (IPNO), Universit\'e Paris-Sud, CNRS-IN2P3, Orsay, France
\item \Idef{org52}Institut f\"{u}r Informatik, Johann Wolfgang Goethe-Universit\"{a}t Frankfurt, Frankfurt, Germany
\item \Idef{org53}Institut f\"{u}r Kernphysik, Johann Wolfgang Goethe-Universit\"{a}t Frankfurt, Frankfurt, Germany
\item \Idef{org54}Institut f\"{u}r Kernphysik, Westf\"{a}lische Wilhelms-Universit\"{a}t M\"{u}nster, M\"{u}nster, Germany
\item \Idef{org55}Institut Pluridisciplinaire Hubert Curien (IPHC), Universit\'{e} de Strasbourg, CNRS-IN2P3, Strasbourg, France
\item \Idef{org56}Institute for Nuclear Research, Academy of Sciences, Moscow, Russia
\item \Idef{org57}Institute for Subatomic Physics of Utrecht University, Utrecht, Netherlands
\item \Idef{org58}Institute for Theoretical and Experimental Physics, Moscow, Russia
\item \Idef{org59}Institute of Experimental Physics, Slovak Academy of Sciences, Ko\v{s}ice, Slovakia
\item \Idef{org60}Institute of Physics, Academy of Sciences of the Czech Republic, Prague, Czech Republic
\item \Idef{org61}Institute of Physics, Bhubaneswar, India
\item \Idef{org62}Institute of Space Science (ISS), Bucharest, Romania
\item \Idef{org63}Instituto de Ciencias Nucleares, Universidad Nacional Aut\'{o}noma de M\'{e}xico, Mexico City, Mexico
\item \Idef{org64}Instituto de F\'{\i}sica, Universidad Nacional Aut\'{o}noma de M\'{e}xico, Mexico City, Mexico
\item \Idef{org65}iThemba LABS, National Research Foundation, Somerset West, South Africa
\item \Idef{org66}Joint Institute for Nuclear Research (JINR), Dubna, Russia
\item \Idef{org67}Konkuk University, Seoul, South Korea
\item \Idef{org68}Korea Institute of Science and Technology Information, Daejeon, South Korea
\item \Idef{org69}KTO Karatay University, Konya, Turkey
\item \Idef{org70}Laboratoire de Physique Corpusculaire (LPC), Clermont Universit\'{e}, Universit\'{e} Blaise Pascal, CNRS--IN2P3, Clermont-Ferrand, France
\item \Idef{org71}Laboratoire de Physique Subatomique et de Cosmologie, Universit\'{e} Grenoble-Alpes, CNRS-IN2P3, Grenoble, France
\item \Idef{org72}Laboratori Nazionali di Frascati, INFN, Frascati, Italy
\item \Idef{org73}Laboratori Nazionali di Legnaro, INFN, Legnaro, Italy
\item \Idef{org74}Lawrence Berkeley National Laboratory, Berkeley, California, United States
\item \Idef{org75}Lawrence Livermore National Laboratory, Livermore, California, United States
\item \Idef{org76}Moscow Engineering Physics Institute, Moscow, Russia
\item \Idef{org77}National Centre for Nuclear Studies, Warsaw, Poland
\item \Idef{org78}National Institute for Physics and Nuclear Engineering, Bucharest, Romania
\item \Idef{org79}National Institute of Science Education and Research, Bhubaneswar, India
\item \Idef{org80}Niels Bohr Institute, University of Copenhagen, Copenhagen, Denmark
\item \Idef{org81}Nikhef, Nationaal instituut voor subatomaire fysica, Amsterdam, Netherlands
\item \Idef{org82}Nuclear Physics Group, STFC Daresbury Laboratory, Daresbury, United Kingdom
\item \Idef{org83}Nuclear Physics Institute, Academy of Sciences of the Czech Republic, \v{R}e\v{z} u Prahy, Czech Republic
\item \Idef{org84}Oak Ridge National Laboratory, Oak Ridge, Tennessee, United States
\item \Idef{org85}Petersburg Nuclear Physics Institute, Gatchina, Russia
\item \Idef{org86}Physics Department, Creighton University, Omaha, Nebraska, United States
\item \Idef{org87}Physics Department, Panjab University, Chandigarh, India
\item \Idef{org88}Physics Department, University of Athens, Athens, Greece
\item \Idef{org89}Physics Department, University of Cape Town, Cape Town, South Africa
\item \Idef{org90}Physics Department, University of Jammu, Jammu, India
\item \Idef{org91}Physics Department, University of Rajasthan, Jaipur, India
\item \Idef{org92}Physik Department, Technische Universit\"{a}t M\"{u}nchen, Munich, Germany
\item \Idef{org93}Physikalisches Institut, Ruprecht-Karls-Universit\"{a}t Heidelberg, Heidelberg, Germany
\item \Idef{org94}Politecnico di Torino, Turin, Italy
\item \Idef{org95}Purdue University, West Lafayette, Indiana, United States
\item \Idef{org96}Pusan National University, Pusan, South Korea
\item \Idef{org97}Research Division and ExtreMe Matter Institute EMMI, GSI Helmholtzzentrum f\"ur Schwerionenforschung, Darmstadt, Germany
\item \Idef{org98}Rudjer Bo\v{s}kovi\'{c} Institute, Zagreb, Croatia
\item \Idef{org99}Russian Federal Nuclear Center (VNIIEF), Sarov, Russia
\item \Idef{org100}Russian Research Centre Kurchatov Institute, Moscow, Russia
\item \Idef{org101}Saha Institute of Nuclear Physics, Kolkata, India
\item \Idef{org102}School of Physics and Astronomy, University of Birmingham, Birmingham, United Kingdom
\item \Idef{org103}Secci\'{o}n F\'{\i}sica, Departamento de Ciencias, Pontificia Universidad Cat\'{o}lica del Per\'{u}, Lima, Peru
\item \Idef{org104}Sezione INFN, Bari, Italy
\item \Idef{org105}Sezione INFN, Bologna, Italy
\item \Idef{org106}Sezione INFN, Cagliari, Italy
\item \Idef{org107}Sezione INFN, Catania, Italy
\item \Idef{org108}Sezione INFN, Padova, Italy
\item \Idef{org109}Sezione INFN, Rome, Italy
\item \Idef{org110}Sezione INFN, Trieste, Italy
\item \Idef{org111}Sezione INFN, Turin, Italy
\item \Idef{org112}SSC IHEP of NRC Kurchatov institute, Protvino, Russia
\item \Idef{org113}SUBATECH, Ecole des Mines de Nantes, Universit\'{e} de Nantes, CNRS-IN2P3, Nantes, France
\item \Idef{org114}Suranaree University of Technology, Nakhon Ratchasima, Thailand
\item \Idef{org115}Technical University of Ko\v{s}ice, Ko\v{s}ice, Slovakia
\item \Idef{org116}Technical University of Split FESB, Split, Croatia
\item \Idef{org117}The Henryk Niewodniczanski Institute of Nuclear Physics, Polish Academy of Sciences, Cracow, Poland
\item \Idef{org118}The University of Texas at Austin, Physics Department, Austin, Texas, USA
\item \Idef{org119}Universidad Aut\'{o}noma de Sinaloa, Culiac\'{a}n, Mexico
\item \Idef{org120}Universidade de S\~{a}o Paulo (USP), S\~{a}o Paulo, Brazil
\item \Idef{org121}Universidade Estadual de Campinas (UNICAMP), Campinas, Brazil
\item \Idef{org122}University of Houston, Houston, Texas, United States
\item \Idef{org123}University of Jyv\"{a}skyl\"{a}, Jyv\"{a}skyl\"{a}, Finland
\item \Idef{org124}University of Liverpool, Liverpool, United Kingdom
\item \Idef{org125}University of Tennessee, Knoxville, Tennessee, United States
\item \Idef{org126}University of the Witwatersrand, Johannesburg, South Africa
\item \Idef{org127}University of Tokyo, Tokyo, Japan
\item \Idef{org128}University of Tsukuba, Tsukuba, Japan
\item \Idef{org129}University of Zagreb, Zagreb, Croatia
\item \Idef{org130}Universit\'{e} de Lyon, Universit\'{e} Lyon 1, CNRS/IN2P3, IPN-Lyon, Villeurbanne, France
\item \Idef{org131}V.~Fock Institute for Physics, St. Petersburg State University, St. Petersburg, Russia
\item \Idef{org132}Variable Energy Cyclotron Centre, Kolkata, India
\item \Idef{org133}Warsaw University of Technology, Warsaw, Poland
\item \Idef{org134}Wayne State University, Detroit, Michigan, United States
\item \Idef{org135}Wigner Research Centre for Physics, Hungarian Academy of Sciences, Budapest, Hungary
\item \Idef{org136}Yale University, New Haven, Connecticut, United States
\item \Idef{org137}Yonsei University, Seoul, South Korea
\item \Idef{org138}Zentrum f\"{u}r Technologietransfer und Telekommunikation (ZTT), Fachhochschule Worms, Worms, Germany
\end{Authlist}
\endgroup

%
%
\end{document}